\documentclass[11pt]{article}

\usepackage[final]{acl}

\usepackage{times}
\usepackage{latexsym}

\usepackage[T1]{fontenc}

\usepackage[utf8]{inputenc}
\usepackage{tabularx}
\usepackage{booktabs}

\usepackage{microtype}

\usepackage{inconsolata}
\usepackage{enumitem}

\usepackage{graphicx}

\usepackage{amsfonts}
\usepackage{amsmath}
\usepackage{amssymb}
\usepackage{multirow} 
\usepackage{subcaption}

\usepackage{listings}
\newcommand{\imnl}{{ICRS}}

\newcommand{\Autoref}[1]{%
  \begingroup
  \def\chapterautorefname{Chapter}%
  \def\sectionautorefname{Sec.}%
  \def\subsectionautorefname{Sec.}%
  \def\subsubsectionautorefname{Sec.}%
  \def\paragraphautorefname{Para.}%
  \def\tableautorefname{Tab.}%
  \def\figureautorefname{Fig.}%
  \def\equationautorefname{Eq.}%
  \def\algorithmautorefname{Alg.}%
  \def\appendixautorefname{App.}%
  \autoref{#1}%
  \endgroup
}

\title{Evaluating Scene-based In-Situ Item Labeling for Immersive Conversational Recommendation}

\author{
  \textbf{Jiazhou Liang\thanks{Equal contribution}\textsuperscript{1}},
  \textbf{Yifan Simon Liu\footnotemark[1]\textsuperscript{1}},
  \textbf{David Guo\footnotemark[1]\textsuperscript{1}}, \\
  \textbf{Minqi Sun\textsuperscript{2}},
  \textbf{Yilun Jiang\textsuperscript{2}},
  \textbf{Scott Sanner\textsuperscript{1,3}}
\\
  \textsuperscript{1}University of Toronto, Toronto, Canada
  \\
  \textsuperscript{2}University of Waterloo, Waterloo, Canada \\
    \textsuperscript{3}Vector Institute of Artificial Intelligence, Toronto, Canada
}

\begin{document}
\maketitle

\begin{abstract}

The growing ubiquity of Extended Reality (XR) is driving Conversational Recommendation Systems (CRS) toward visually immersive experiences.   We formalize this paradigm as \emph{Immersive CRS} (\imnl{}), where recommended items are highlighted directly in the user’s scene-based visual environment and augmented with in-situ labels. While item recommendation has been widely studied, the problem of how to select and evaluate which information to present as immersive labels remains an open problem.  
To this end, we introduce a principled categorization of information needs into \emph{explicit intent satisfaction} and \emph{proactive information needs} and use these to define novel evaluation metrics for item label selection. We benchmark IR-, LLM-, and VLM-based methods across three datasets and \imnl{} scenarios: fashion, movie recommendation, and retail shopping.
Our evaluation reveals three important limitations of existing methods: (1) they fail to leverage scenario-specific information modalities (e.g., visual cues for fashion, metadata for retail), (2) they present redundant information that is visually inferable, and (3) they poorly anticipate users’ proactive information needs from explicit dialogue alone.  In summary, this work provides both a novel evaluation paradigm for in-situ item labeling in \imnl{} and highlights key challenges for future work.

\end{abstract}
\section{Introduction}

\begin{figure}
    \centering
    \includegraphics[width=0.97\linewidth]{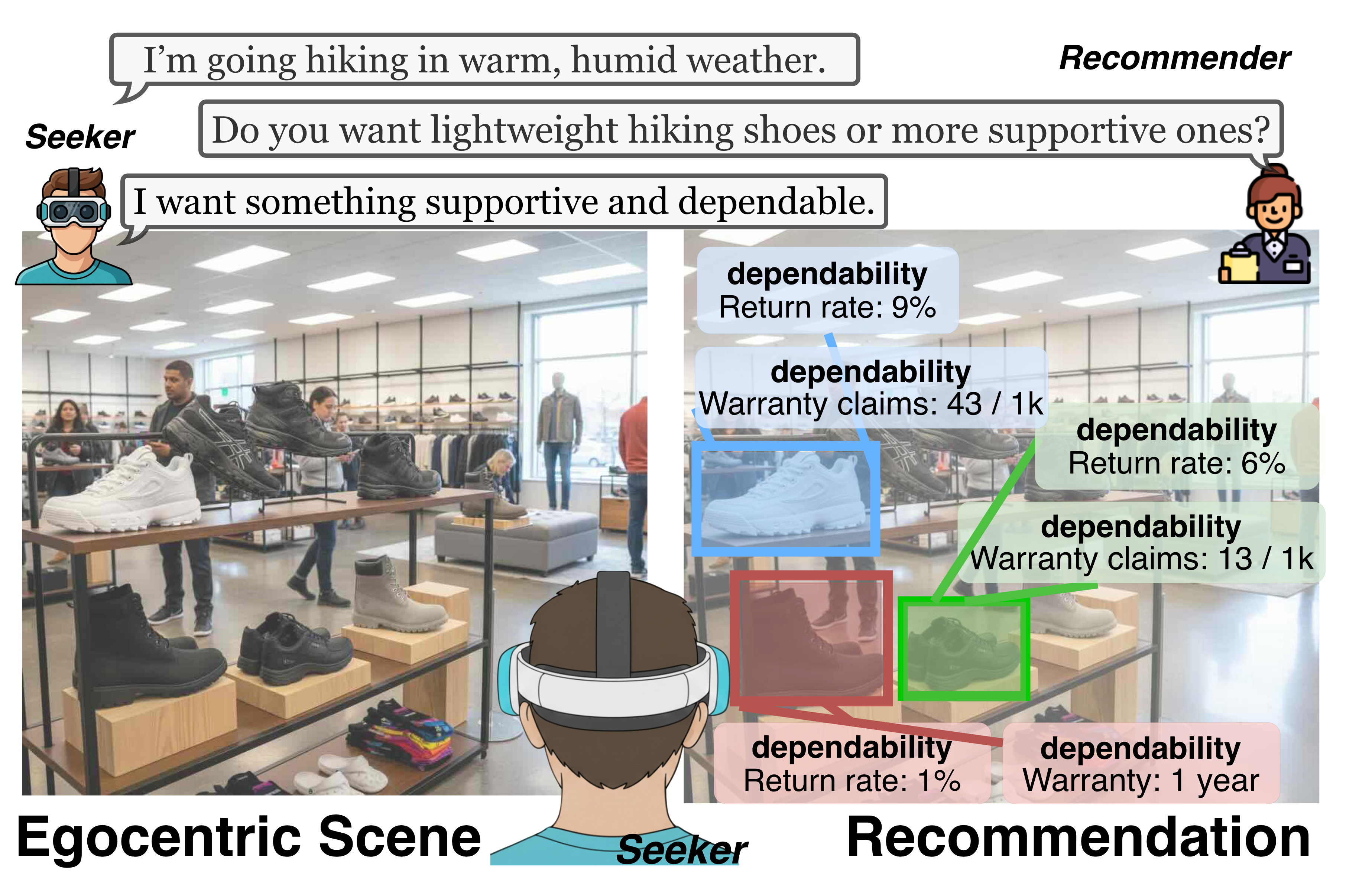}
    \caption{ \textbf{Left}:  Seeker’s egocentric view of the scene captured by the immersive system. \textbf{Right}: Recommended items are highlighted and augmented with in-situ immersive labels within the scene.
    }
    \label{fig:example}
\end{figure}

The growing ubiquity of Extended Reality (XR) has driven rapid adoption across a wide range of domains \citep{mendoza2023arsurvey}. Such technology enables \emph{immersive} systems with input of egocentric scenes (e.g., head-mounted cameras) (cf. left of \Autoref{fig:example}) and information delivery via situated visualization~\citep{lee2023design} (e.g., \emph{in-situ labels}) within the user’s Point of View (POV).

However, immersive interfaces remain largely unexplored in the existing Conversational Recommendation Systems (CRS) literature. Motivated by this gap, we introduce Immersive CRS (\imnl{}), a new problem that transforms CRS from language-focused interactions to egocentric, in-situ experiences. CRS recommends items from a global catalog in response to seeker requests expressed through natural-language (NL) utterances, modeling seeker needs and preferences as inferred from the conversational context \citep{li2018redial}. In contrast, \imnl{} recommends items identified from the scene and highlights them directly within the seeker’s POV (cf. right of \Autoref{fig:example}). Each recommended item is accompanied by \emph{immersive labels}, a virtual element that is presented in-situ to referent items, conveying information that is not visually accessible from the physical items (\emph{latent}) to resolve the seeker’s uncertainty and information needs.



\begin{table*}[t]
\centering
\caption{Comparison between CRS, Multimodal CRS, and the proposed ICRS across four key dimensions.}
\label{tab:rw_novelty}
\scriptsize
\renewcommand{\arraystretch}{1.4} 

{%
\begin{tabularx}{\linewidth}{
    >{\raggedright\arraybackslash\hsize=0.5\hsize}X |
    >{\raggedright\arraybackslash\hsize=0.3\hsize}X |
    >{\raggedright\arraybackslash\hsize=1\hsize}X |
    >{\raggedright\arraybackslash\hsize=2.5\hsize}X 
}

& \textbf{CRS} 
& \textbf{Multimodal CRS} 
& \textbf{Immersive CRS} \\
\specialrule{.2em}{.1em}{.1em}   

\textbf{Input} 
& \multicolumn{2}{c|}{Multi-turn conversation between seeker and recommender}
& Conversation + \textbf{Egocentric Scene from Seeker's POV} \\
\hline 

\textbf{Candidate Items} 
& \multicolumn{2}{c|}{A set of predefined items in the large global catalog}
& Items are identified in the \textbf{egocentric scene}.\\
\hline 

\textbf{Item Attributes} 
&
\multicolumn{1}{c|}{Textual attributes (e.g., metadata, review)}
& Textual + Predefined multimodal attributes
& \textbf{Visual segments from egocentric scene} +   \textbf{external attributes} identified via item mapping from global corpus \\
\hline 

\hline 

\textbf{Output} 
& \multicolumn{2}{c|}{Recommended items encapsulated within multi-turn utterances}
& Recommended items are highlighted in the egocentric scene with \emph{immersive labels} to resolve seeker information needs.  \\
\specialrule{.2em}{.1em}{.1em}
\end{tabularx}%
}
\end{table*}

We illustrate the end-to-end pipeline of \imnl{} in \Autoref{fig:framework}. While upstream components such as item segmentation and recommendation have been extensively studied~\citep{he2023large, kirillov2023segment_anything}, determining and evaluating information in immersive labels, an essential component of \imnl{} for fully leveraging the capabilities of immersive interfaces, remains largely unexplored. We argue that immersive labels should convey information that (1) justifies why an item aligns with the seeker’s explicit intent as expressed in the current conversation (\emph{Explicit Intent Satisfaction}), and (2) proactively provides decision-relevant information before it is explicitly requested (\emph{Information Needs}). 

Accordingly, we propose evaluation criteria and metrics for each consideration. We adapt existing CRS datasets with novel, manually curated ground-truth annotations that are specifically designed for immersive label selection and conduct extensive experiments across contemporary methods with IR, LLM, and VLM capabilities in three scenarios: Fashion, Movie, and Retail. Our evaluation results reveal three key limitations of existing methods:
\begin{enumerate}
    \item They fail to leverage visual and textual information germane to the scenario (e.g., visual cues are more informative in Fashion, while textual metadata are more relevant in Retail).
    \item They often present information that is already visually inferable (e.g., color), which offers marginal value w.r.t.\ users’ information needs.
    \item They struggle to preemptively anticipate information that users will need, but have not yet explicitly requested for the recommended item.
\end{enumerate}

Together, our findings position label selection for \imnl{} that fully leverage immersive interfaces as a challenging yet underexplored research problem.

\section{Related Work}






\paragraph{CRS and Multimodal CRS}
CRS infer user preferences from multi-turn dialogue and recommend items predefined in a global catalog~\citep{li2018redial, hayati2020inspired,  he2023large, alessio2025cosrec, eqr2024, eqr2025, liu2025multimodal}, with multimodal variants incorporating item images and metadata to enhance reasoning~\citep{mg_shopdial, wang2025muse, guo2025vogue} or incorporating a VLM conversational agent~\citep {liu2026semantic} as the recommender. As shown in \Autoref{tab:rw_novelty}, \imnl{} defines a distinct paradigm for immersive interfaces with egocentric, in-situ interaction that is not addressed by prior work.


\paragraph{Context-Aware Conversation.}
Prior work on context-aware conversation largely addresses referential grounding and location-aware utterance generation, as seen in SIMMC and spatial AR dialogue systems~\citep{kottur2021simmc2, wu-etal-2023-simmc, long2023spring, lee2024gazepointar}. While effective for linking dialogue to visual or spatial context, these approaches are not designed for preference-oriented recommendation and do not define what item information should be shown via item labels to support recommendation interactions.


\paragraph{Situated Visualization}\label{sec:label_rendar}
In situated visualization~\citep{lee2023design}, prior research has addressed the placement and management of visualized information \textit{in-situ} within the seeker’s POV~\citep{bressa2021s}. However, these works primarily focus on layout and spatial alignment, rather than the retrieval criteria of information in the label.

\begin{figure*}
    \centering
        \includegraphics[width=\linewidth]{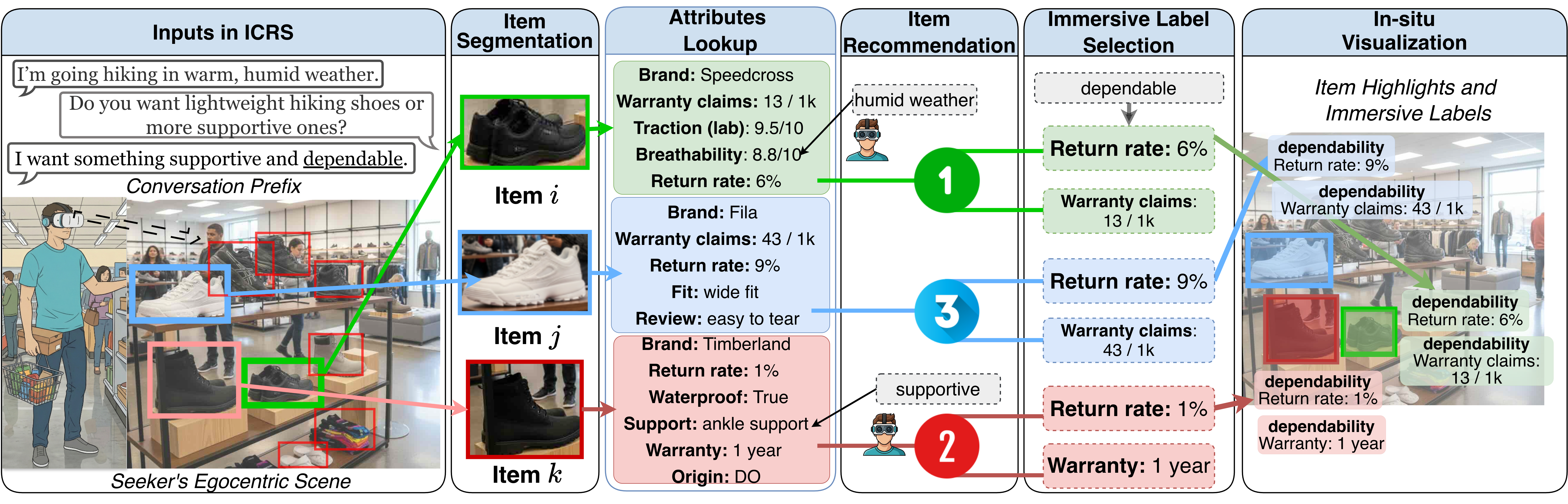}
    \caption{\textbf{Components in \imnl{}}. Given a conversation prefix and seeker's egocentric scene (left), candidate items in \imnl{} are identified via segmentation and enriched with external attributes through visual lookup. \imnl{} then ranks candidate items and selects in-situ labels (center) that address the seeker’s information needs (e.g., dependability). Recommended items and immersive labels are highlighted and augmented in the scene (right). }
    \label{fig:framework}
\end{figure*}

\paragraph{Image Identification and Attribute Lookup}\label{images_seg}
Prior work leverages foundation models like Segment Anything (SAM)~\citep{kirillov2023segment_anything} with vision–language pretraining (e.g., CLIP~\citep{radford2021clip} to identify textually related segments or reverse lookup external attributes from product segmentation in environments (shelf recognition)~\citep{melek2024groceryreview,pettersson2024multimodal}.  These methods provide the raw inputs (cf. \Autoref{fig:framework}) used by the \imnl{} labeling approaches we evaluate.
\section{Immersive CRS}
\label{sec:framework} 
\label{sec:crs_prelim}

\paragraph{Preliminary Definition} The Immersive CRS (\imnl{}) adapts the CRS definition~\citep{he2023large}. Let $\mathcal{U}$ represent two sides of the conversation: a \emph{seeker} and a \emph{recommender}.

Given global candidate item set $\mathcal{I}^{\text{glb}}$, a conversation can be denoted as a sequence $\mathcal{C} = \{(u_t, s_t, \mathcal{I}^{\text{glb}}_t)\}_{t=1}^{T}$. At each turn $t$, either seeker or recommender generates an utterance $s_t = (\text{w}_i)_{i=1}^{m}$, with each word $\text{w}_i$ belonging to the vocabulary, and optionally reference items $\mathcal{I}_t^{\text{glb}} \subseteq \mathcal{I}^{\text{glb}}$. 

Given a conversation prefix $\mathcal{C}_{1:t}$ (i.e., the conversation context before turn $t$), the CRS (1) infers the seeker's evolving preferences and (2) generates an appropriate utterance (e.g., item recommendation).

\paragraph{Overview}
 As summarized in \Autoref{tab:rw_novelty}, \imnl{} differs from conventional CRS along three dimensions: (1)~in addition to conversational context, the system incorporates the seeker egocentric view of the scene as input; (2) the candidate item set is constrained to items physically present in scene $\mathcal{I}^{\text{env}}$ rather than drawn from a global catalog; and (3) recommendations are delivered via visual highlighting from the seeker's POV, each augmented with immersive labels to address seeker information needs. 

Presenting immersive labels that convey effective information in the seeker's POV requires resolving a sequence of interrelated decisions: identifying items and associated attributes from the egocentric scene, determining which items should be recommended given the conversation prefix, retrieving information from attributes, and rendering this information in situ as immersive labels. As illustrated in \Autoref{fig:framework}, we structure \imnl{} into:
\begin{itemize}
\item \textbf{Item Segmentation.} Segmenting objects in the egocentric scene to map physical referents to the scene-based candidate item set, where segments serve as the visual information.

    \item \textbf{Attributes Lookup.}  
    Identifying external attributes from each item's visual information.

    \item \textbf{Item Recommendation.}  
    Selecting a subset of items that best match the seeker’s preferences as expressed in the conversation.

    \item \textbf{Immersive Label Selection.}  
    Choosing attributes of recommended items that satisfy the user's explicit and proactive information needs.

    \item \textbf{In-situ Visualization.}  
Highlighting recommended items and displaying their associated in-situ labels in the seeker’s POV.
\end{itemize}
\noindent We next discuss each stage of \imnl{} in detail.
\subsection{Segmentation and Attributes Lookup}\label{sec:catalog_enrich}
\label{sec:image_seg}
Given an egocentric view of the scene $\mathcal{E}$ captured by an immersive system, an image segmentation model extracts a set of visual segments
$
\mathcal{V} = \{ \mathcal{V}^{(1)}, \mathcal{V}^{(2)}, \ldots, \mathcal{V}^{(N)} \}.
$
The physical referent in the each segment $\mathcal{V}^{(i)} \subseteq \mathcal{V}$ is defined as a candidate item $i$ in $\mathcal{I}^{\text{env}}$. And $\mathcal{V}^{(i)}$ represents the visual appearance of this item.

We assume that each item $i \in \mathcal{I}^{\text{env}}$ is associated with \emph{latent} attributes not directly inferable from its physical appearance (e.g., reviews or sales)
which provides a pool of candidate attributes for immersive labels.
Reverse-lookup models map each $\mathcal{V}^{(i)}$ to an entity in a corpus $\mathcal{L}$ (e.g., a warehouse inventory system) that contains the uniquely identifiable attribute for each item $\mathcal{L}^{(i)}$.

Image segmentation and reverse-lookup have established solutions in the literature (cf. \Autoref{images_seg}). Thus, they are not the primary focus of this work.



\subsection{Item Recommendation}
\label{sec:item_rec}


Given a conversation prefix $\mathcal{C}_{1:t}$, the objective of item recommendation, consistent with CRS, is to select items $\mathcal{I}^{\text{env}}_t \subseteq \mathcal{I}^{\text{env}}$ that best align with the seeker’s preferences. Although item recommendation has been widely studied in CRS \cite{li2018redial, hayati2020inspired, he2023large}, it plays a key role in \imnl{} by determining which items are highlighted and thus receive immersive labels.
\paragraph{Evaluation.}
We adopt a standard offline top-$K$ evaluation protocol \citep{cremonesi2010topn} and assume a ground truth relevant item set $\mathcal{I}^{\text{env}*}_t \subseteq \mathcal{I}^{\text{env}}$ for the conversation prefix $\mathcal{C}_{1:t}$.

In \imnl{}, recommendations are delivered via visual highlighting on head-worn XR devices. Highlighting irrelevant items can distract seekers and occlude relevant content, given the limited field of view, necessitating highly precise recommendations. Because items are spatially highlighted instead of presented as a ranked list, we do not consider relative ordering. Thus, we adopt
\noindent\texttt{Precision@K}
$
=
\frac{1}{K}
\sum_{k=1}^{K}
 \mathbb{I}\!\left[
 \hat{i}^{\text{env}}_{t,k}
\in
\mathcal{I}^{\text{env}*}_t
\right]
$ where $\hat{i}^{\text{env}}_{t,k}$ denotes the item ranked at position $k$ at turn $t$ and $\mathbb{I}$ is the 0-1 indicator.

Prior XR studies also show that high overlay density degrades visual search and comprehension, motivating interfaces that limit the number of in-situ labels (typically $\leq 10$) \citep{trepkowski2019fovdensity,lin2021labeling}. As each item may possess multiple latent attributes, this clutter constraint is further amplified in \imnl{}. We adopt $K \leq 3$ as a practical perceptual capacity threshold for highlighted items.


\subsection{Immersive Label Selection}
\label{sec:item_label}
Given recommended items $\mathcal{I}^{\text{env}}_t$, let
$\mathcal{A}^{(i)} = \{a_1^{(i)}, \ldots, a_m^{(i)}\}$ be an atomic attribute set for each item $i$ in $\mathcal{I}^{\text{env}}_t$, where each 
$a_j^{(i)}$ is a self-contained statement decomposed from the full attribute set 
$\mathcal{L}^{(i)}$. $\mathcal{A}^{(i)}$ avoids composite information in $\mathcal{L}^{(i)}$ (e.g., mixed positive and negative aspects within a review) (cf. \Autoref{app:dataset_adaptation} for examples of decomposition in real datasets). Immersive label selection aims to determine a subset of attributes
$\mathcal{A}^{(i)}_t \subseteq \mathcal{A}^{(i)}$ as immersive labels for each item 
$i \in \mathcal{I}^{\text{env}}_t$.


We posit that immersive labels should address two levels of user intent: (1) \textit{explicit intent} derived from explicitly stated preferences in the conversation, and (2) \textit{latent proactive} information needs that are not expressed but can be inferred from the conversation. The former justifies recommended items, which are an essential requisite for CRS, while the latter addresses the proactive surfacing of decision-critical, non-visual item information, reducing reliance on back-and-forth clarification.

The ``latent proactive'' level further entails two novel reasoning challenges compared to CRS: What information should be selected that complements the seeker's direct observations within the egocentric scene? (\emph{latent awareness}) And what information should be presented \emph{before} the seeker explicitly requests it, based on intent inferred from the dialogue? (\emph{proactive disclosure}) 

\paragraph{Criteria.}
\label{par: label_criteria}
To assess how well systems meet these requirements and address the above challenges, we propose the following criteria:

\noindent\underline{\textit{Seeker Retroactive Information Needs (RIN)}} evaluates the ability to provide surface-level justification for the recommended item by measuring how well the selected atomic attributes address the seeker’s explicitly stated requests and preferences in  $\mathcal{C}_{1:t}$.

\noindent\underline{\textit{Seeker Proactive Information Need (PIN)}} evaluates a system's ability to infer the seeker's latent information needs that have not been explicitly stated in previous conversation $\mathcal{C}_{1:t}$.

\noindent\underline{\textit{Expert Inferred Information Need (EIN)}} evaluates a system's ability to surface decision-relevant information by measuring alignment with information that a knowledgeable recommender can infer about the seeker's needs, information the seeker may not know to ask about.


\begin{figure}[!h]
    \centering
    \includegraphics[width=\linewidth]{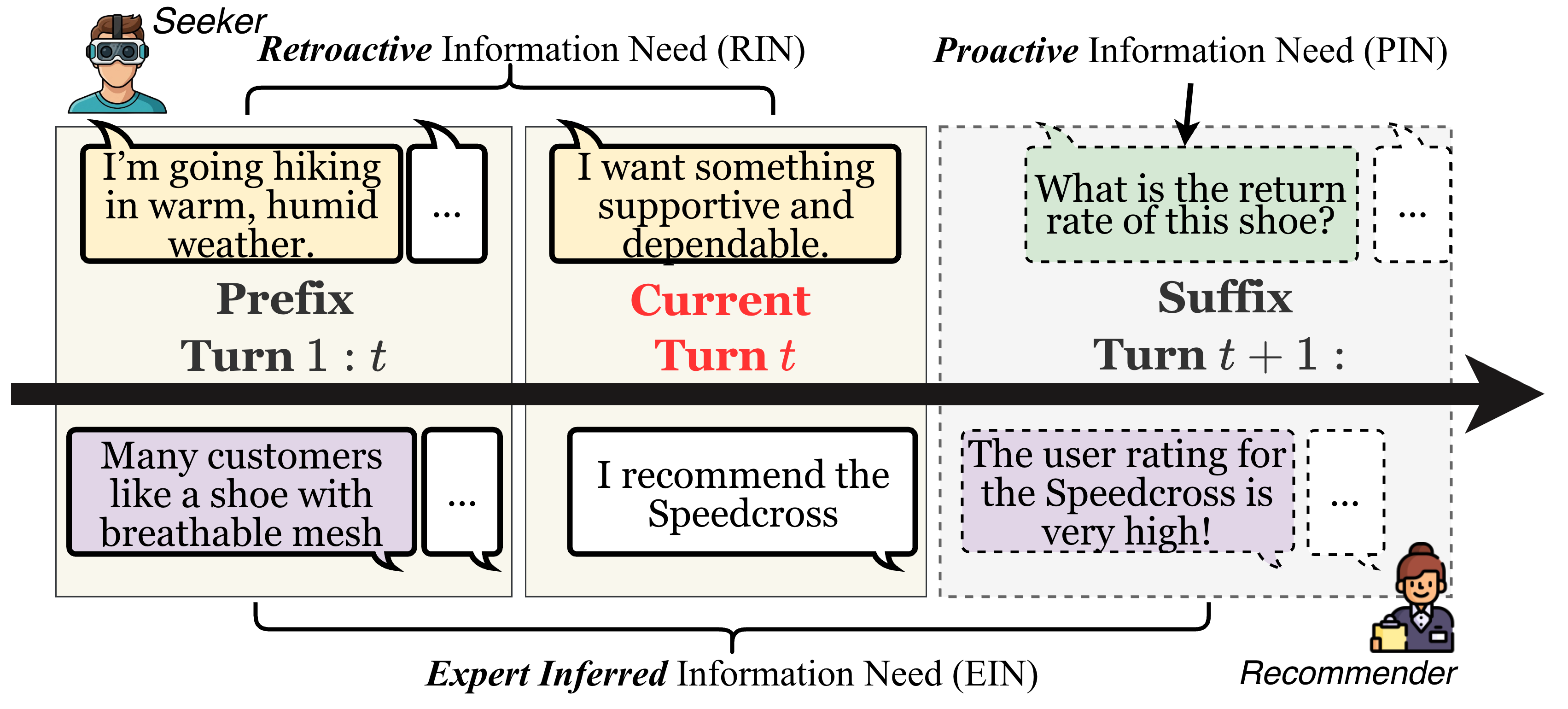}
\caption{\textbf{Evaluation criteria for immersive label selection.} (top left) RIN addresses explicitly stated seeker preferences in $\mathcal{C}_{1:t}$; (top right) PIN captures proactive seeker requests in $\mathcal{C}_{t+1:T}$; (bottom) EIN reflects the seeker's information needs inferred by the recommender. As $t$ advances and more information is revealed through the conversation, the set of retroactive/proactive needs changes accordingly.}
    \label{fig:information_needs}
\end{figure}

As illustrated in Fig.~\ref{fig:information_needs}, \textit{RIN} assesses whether a system can surface attributes that directly address the seeker's explicitly stated requests in $\mathcal{C}_{1:t}$. \textit{PIN} is more challenging, as it requires reasoning over latent preferences and sources of uncertainty in recommended items that are not directly stated in $\mathcal{C}_{1:t}$.
\textit{RIN} and \textit{PIN} are inherently dynamic with respect to the current turn $t$. As the conversation progresses, previously proactive requests become explicitly stated, shifting from \textit{PIN} into \textit{RIN}. Consequently, an effective immersive label selection must be sensitive not only to what has been stated, but also to what remains latent at each turn.

\textit{EIN}, on the other hand, captures knowledge that a knowledgeable recommender surfaces based on domain experience and prior interactions with similar recommendation situations. For instance, an expert recommender may proactively highlight warranty coverage, latent functionality, or commonly requested item attributes, information the seeker may not know to ask about, yet is decision-critical for resolving item uncertainty.


\paragraph{Ground-truth.}
\label{par: label_gt}
We define practical guidelines for constructing the ground-truth immersive label set 
$\mathcal{A}^{(i)*,\text{crit}}_t \subseteq \mathcal{A}^{(i)}$ for
$\text{crit} \in \{\textit{RIN}, \textit{PIN}, \textit{EIN}\}$. 
We contend that any full conversation contains different forms of seeker and recommender intent relevant to all three criteria. For \textit{RIN} and \textit{PIN}, seekers naturally express their information needs through seeker-side requests across the conversation. For \textit{EIN}, helpful domain expert-guided information naturally surfaces in recommender-side utterances~\citep{kostric2025know}.

Accordingly, we introduce three utterance-level intent tags: \textit{retroactive user request}, \textit{proactive user request}, and \textit{expert explanation}, each aligned with a criterion. Specifically, under offline evaluation, the full conversation $\mathcal{C}_{1:T}$ is divided into a system observable prefix $\mathcal{C}_{1:t}$ and an unobserved suffix $\mathcal{C}_{t:T}$. Each utterance is annotated with zero or more intent tags:
\begin{enumerate}
    \item \textit{Retroactive Seeker Request:} seeker utterances in $\mathcal{C}_{1:t}$ that express explicit preferences, for which addressing them satisfies \textit{RIN}.
    \item \textit{Proactive Seeker Request:} seeker utterances in $\mathcal{C}_{t:T}$ that reveal information needs not stated before turn $t$ (e.g., follow-up questions), for which addressing them satisfies \textit{PIN}.
    \item \textit{Expert Explanation:} recommender utterances in $\mathcal{C}_{t:T}$ that provide justifications for recommendation or inference on seeker's info-needs, for which being relevant to them satisfies \textit{EIN}.
\end{enumerate}

\begin{figure}[!h]
    \centering
    \includegraphics[width=\linewidth]{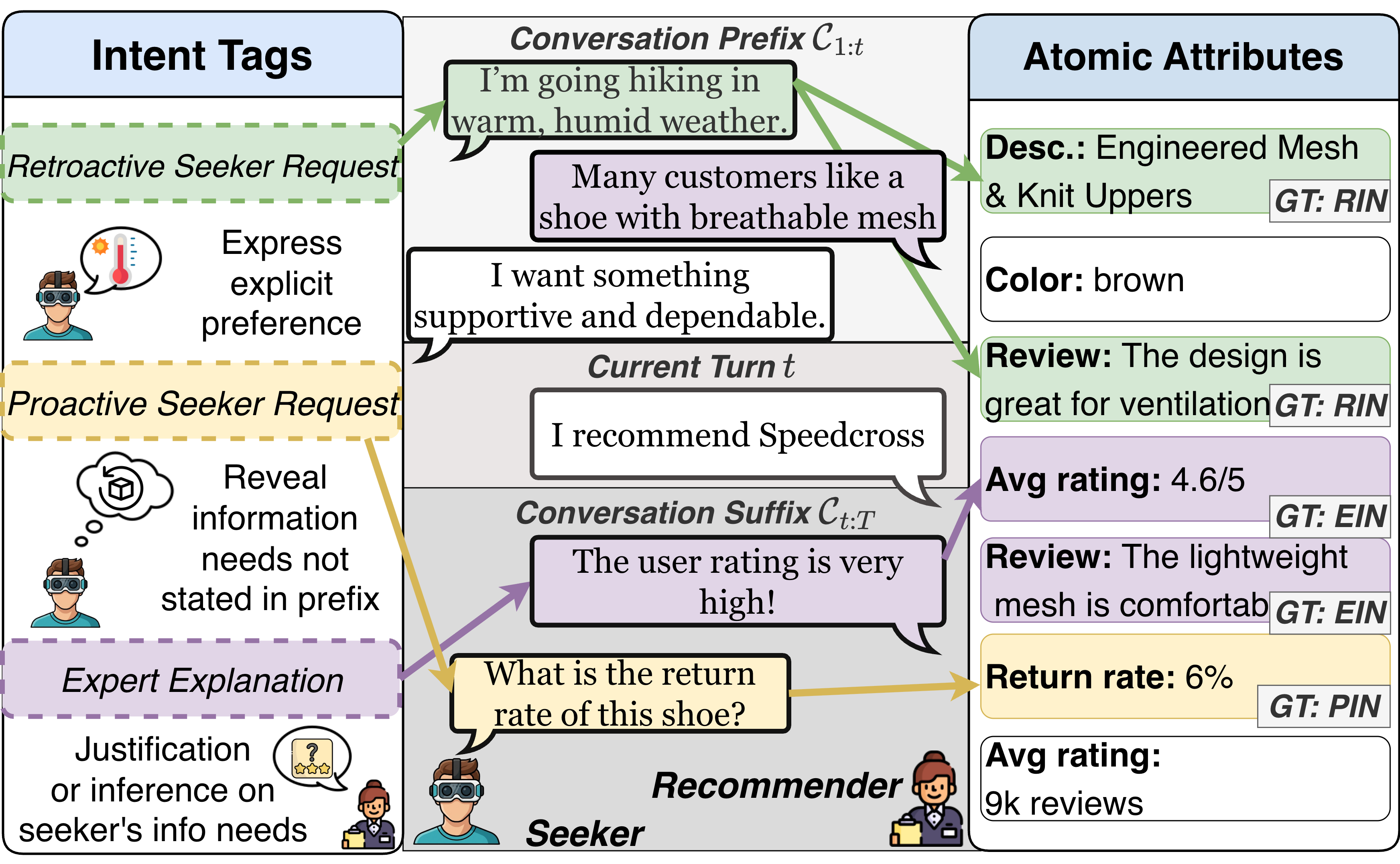}
\caption{\textbf{Ground Truth for Label Selection.}
Ground truth for the three proposed criteria is constructed by assigning intent tags (left) to each utterance in the full conversation (center), and selecting atomic attributes that satisfy at least one utterance (right) as ground truth.}
    \label{fig:label_selections}
\end{figure}

We illustrate an exemplary utterance for each intent tag in \Autoref{fig:label_selections}. This produces a set of utterances for each criterion based on the corresponding intent tag.  $a^{(i)}_k \in \mathcal{A}^{(i)}$ is considered relevant to a criterion if it semantically addresses at least one utterance in its corresponding set.
To enforce the latent requirement, $a^{(i)}_k$ must also not be 
inferable from the item's segmentation $\mathcal{V}^{(i)}$.  
Thus, each item $i$ has three \underline{crit}erion-specific ground-truth sets
$\mathcal{A}^{(i)*, {\text{crit}}}_t$ . For a detailed description of the ground-truth annotation process, we refer the reader to App.~\ref{app:ground_truth}.




\paragraph{Evaluation.}
\label{par: label_eval}

We evaluate \emph{Immersive Label Selection} using \texttt{Precision@K} ($K \leq 3$), enforcing $10$ or fewer labels to prevent clutter within the augmented egocentric scene. We report $\texttt{mP@K}_{\text{crit}}$ (Mean Per-Item Precision at $K$ for \texttt{crit}):
\begin{equation*}
\texttt{mP@K}_{\text{crit}}
=
\frac{1}{|\mathcal{I}_{t}^{\text{env}}|K}
\sum_{i \in \mathcal{I}_{t}^{\text{env}}}
\sum_{k=1}^{K}
\mathbb{I}\!\bigl[
\hat{a}^{(i), \text{crit}}_{t,k} \in \mathcal{A}^{(i)*, \text{crit}}_t
\bigr]
\end{equation*}
\noindent
$\hat{a}^{(i),\text{crit}}_{t,k}$ denotes the $k$-th ranked attribute for item $i$
under criterion $\text{crit}$ at turn $t$; $\mathbb{I}$ is the 0-1 indicator.
\subsection{In-Situ Visualization}
Each recommendation $i \in \mathcal{I}_{t}^{\text{env}}$ is visually anchored (e.g., via bounding boxes), with its selected attributes $\mathcal{A}_t^{(i)}$ as \emph{immersive labels}.   
The technical implementation of this component 
is well-explored in the situated visualization literature (cf. \Autoref{images_seg}).



\section{Datasets}
We evaluate on three datasets and scenarios where immersive labels can be highly beneficial:
\begin{itemize}
\item \textbf{Fashion:}
In fashion shopping, decision-making often relies on a fusion of visual cues and non-visual information (e.g., reviews, ratings, functionality). We use VOGUE \citep{guo2025vogue}, which contains fashion shopping dialogues with curated product images and rich item metadata.

\item\textbf{Retail:}
In consumer-product settings, visually similar items require non-visual attributes for effective comparison. We use CoSRec \citep{alessio2025cosrec}, which links dialogues to metadata and images from the Amazon Product Search \citep{mcauley2015amazon, he2016amazon}.

\item\textbf{Movie:}
In the cinema scenario, visual information (e.g. posters, advertisements etc.) alone is insufficient. We use ReDial \citep{li2018redial} and INSPIRED \citep{hayati2020inspired}, augmented with posters, metadata, and reviews in~\citet{tmdb_2026}.
\end{itemize}

\begin{figure}
  \centering
    \begin{subfigure}[t]{0.32\linewidth}
    \centering
    \includegraphics[width=\linewidth]{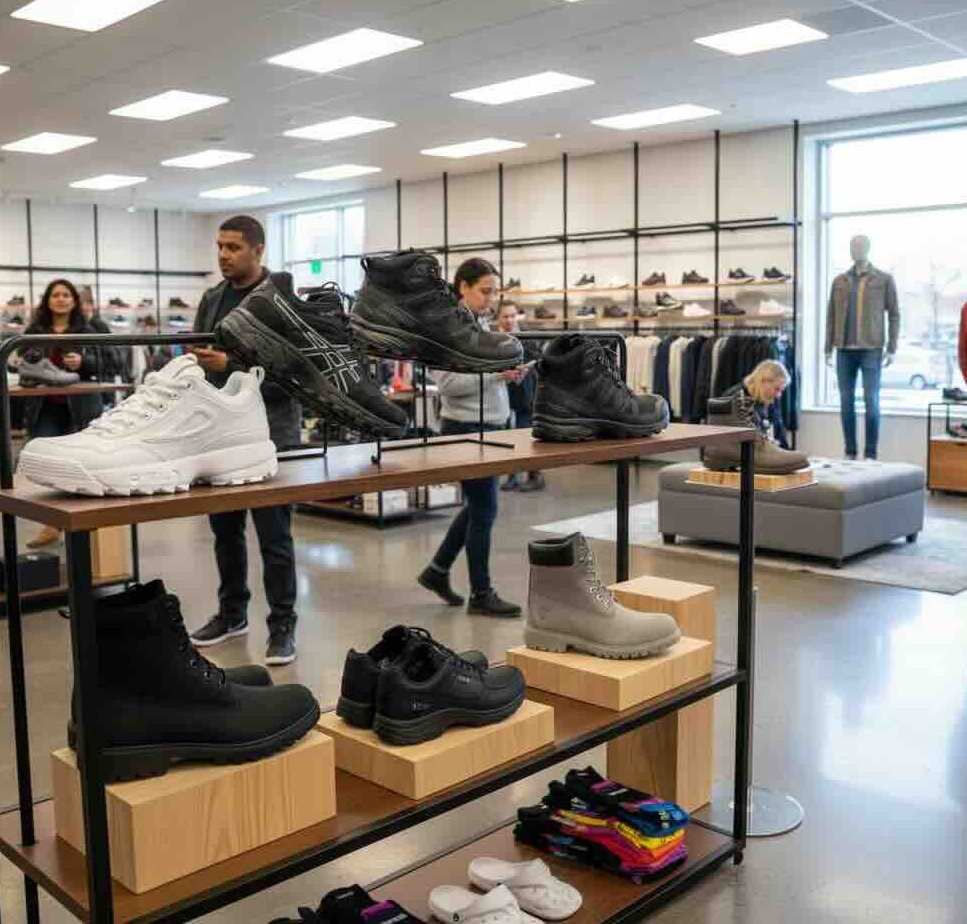}
    \caption{Fashion}
  \end{subfigure}
  \begin{subfigure}[t]{0.32\linewidth}
    \centering
    \includegraphics[width=\linewidth]{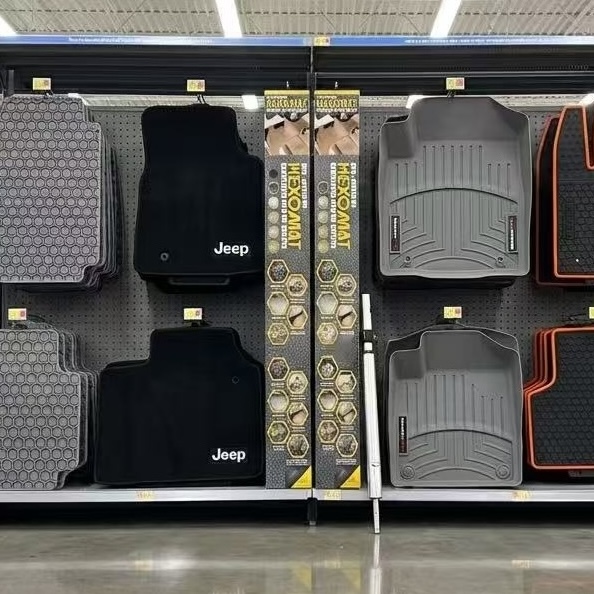}
    \caption{Retail}
  \end{subfigure}
  \begin{subfigure}[t]{0.32\linewidth}
    \centering
    \includegraphics[width=\linewidth]{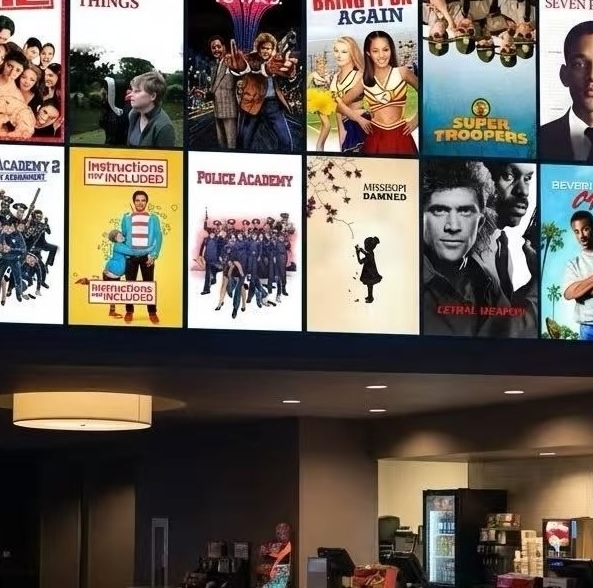}
    \caption{Movie}
  \end{subfigure}

  \caption{The illustration of the egocentric scene $\mathcal{E}$ used in each of the three selected scenarios.}
  \label{fig:three-plots}
\end{figure}

\paragraph{Dataset Adaptation}
\label{sec:dataset_adaptation} 
Existing CRS datasets are not directly applicable to \imnl{} evaluation due to fundamental setting differences (cf. \Autoref{tab:rw_novelty}). Therefore, we introduce a unified pipeline to curate \imnl{} datasets from existing CRS datasets (cf. \Autoref{app:dataset_adaptation} for full description).  We retain original conversations and
candidate items as $\mathcal{C}_{1:t}$ and $\mathcal{I}^{\text{env}}$.
For each $\mathcal{C}_{1:t}$, we synthesize an egocentric scene that reflects how items in $\mathcal{I}^{\text{env}}$ would appear in immersive settings (cf. \Autoref{fig:three-plots} for example). Ground-truth $\mathcal{I}^{\text{env}*}_t$ are inherited from the original datasets. Ground-truth $\mathcal{A}^{(i)*, {\text{crit}}}_t$ are curated by judging each attribute in $\mathcal{A}^{(i)}$ with intent-tagged utterances for \texttt{crit}.\footnote{All experimental results and curated datasets are available at \href{https://github.com/D3Mlab/ICRS}{https://github.com/D3Mlab/ICRS}.}

\section{Experimental Results}
Across the three scenarios, we evaluate the performance of existing methods adapted for \imnl{}. As segmentation and attribute lookup are well studied in prior work, we assume perfect segmentation and item–attribute mappings, and focus on Item Recommendation and immersive label selection. 

Conversation prefixes $\mathcal{C}_{1:t}$  are utterances preceding the first ground-truth recommendation to avoid information leakage from explicitly mentioned ground-truth items and subsequent information requests (cf. \Autoref{app:conversation_prefix} for comparisons with using the full $\mathcal{C}$).
We summarize the key findings due to space and defer the full results to \Autoref{app:full_result}.

\subsection{Item Recommendation}\label{sec:item_rec_exp}

\paragraph{Methods.} We evaluate the following item recommendation methods adapted to \imnl{}:

\noindent \underline{\textit{Retrieval-Based CRS.}}
\begin{itemize}
    \item \emph{Lexical Matching} \citep{robertson2009bm25} ranks items by lexical overlap in $\{\mathcal{C}_{1:t},\mathcal{L}_i\}_{i=1}^{\mathcal{I}^{\text{env}}}$.

\item  \emph{Dense Retrieval} \citep{karpukhin-etal-2020-dense} embeds $\mathcal{C}_{1:t}$, and $\{\mathcal{L}_i\}_{i=1}^{\mathcal{I}^{\text{env}}}$ into a shared semantic space to select $i$ with the shortest in-between the choice of geometric distance~\citep{liu2025ma}.

\item \emph{Re-ranking} \citep{nogueira-etal-2020-document} re-orders Lexical Matching by scoring the top-$K$ items via a cross-encoder that jointly encoding \{$\mathcal{C}_{1:t}, \hat{\mathcal{I}}^{\text{env}(k)}_\text{t|Lexical} \}_{k=1}^K$.
\end{itemize}

\noindent \underline{\textit{VLM as zero-shot CRS}}~\citep{he2023large} \noindent 
\begin{itemize}
\item \emph{Pointwise}~\citep{tian2024mmrec} VLM independently scores each item $i$ in $\mathcal{I}^{\text{env}}$ conditioned on the conversation $\mathcal{C}_{1:t}$, by prompting VLM with task description $\mathcal{S}$, with item information in either $\mathcal{L}^{(i)}$, $\mathcal{V}^{(i)}$, or both. 

\item  \noindent \emph{Listwise} VLM instead directly ranks the full $\mathcal{I}^{\text{env}}$. This enables joint comparison to better utilize the shared information.
\end{itemize}

\paragraph{Experiment Setup.} For Dense retrieval, we use \textsc{Qwen3-8B} as the embedder~\citep{qwen3_technical_report_2025}. 
For VLM-based approaches, we consider three representative backbones: \textsc{GPT-5.1}~\citep{openai_gpt5_1_2025}, 
\textsc{Gemini-2.5-Pro}~\citep{comanici2025gemini25}, and \textsc{Qwen3-VL-30B-Instruct}~\citep{bai2025qwen25vl}.

\paragraph{VLM vs. Retrieval-based CRS.}

\begin{figure}[!h]
    \centering
    \includegraphics[width=\linewidth]{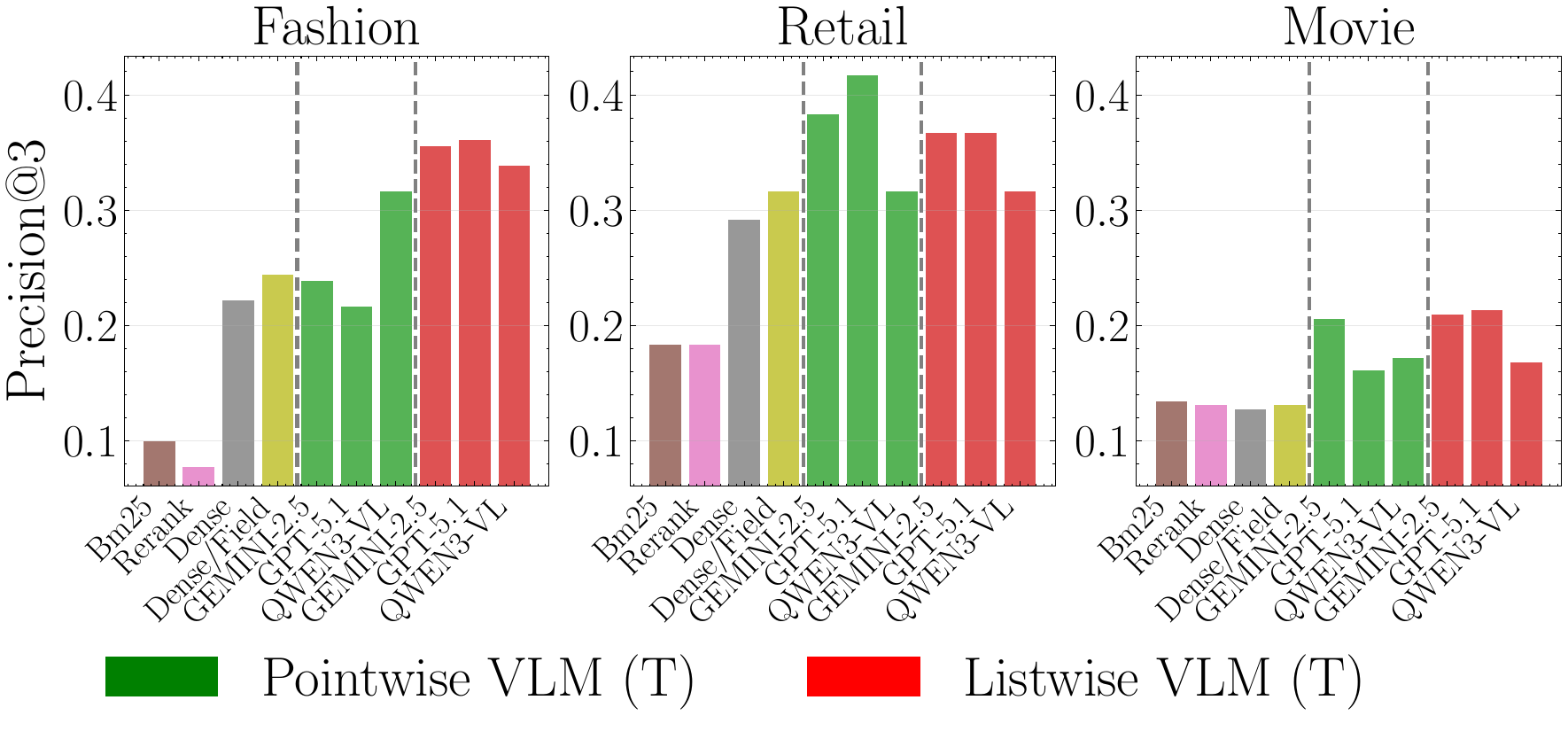}
\caption{\texttt{Precision@3} of retrieval-based and zero-shot VLM-based CRS using textual item attributes only (T). Across all three scenarios, VLM-based methods consistently outperform retrieval-based baselines.}
    \label{fig:llm_vs_retrial}
\end{figure}

We compare two dominant CRS paradigms: retrieval-based and zero-shot VLM-based. For a fair comparison, both approaches rely exclusively on textual item attributes (cf. \Autoref{fig:llm_vs_retrial}). Across all three scenarios, VLM-based methods consistently outperform retrieval-based baselines. This trend reflects the stronger reasoning capacity of frontier VLM backbones in modeling conversational intent.

\paragraph{Item Modality in CRS.}

\begin{figure}
    \centering
    \includegraphics[width=\linewidth]{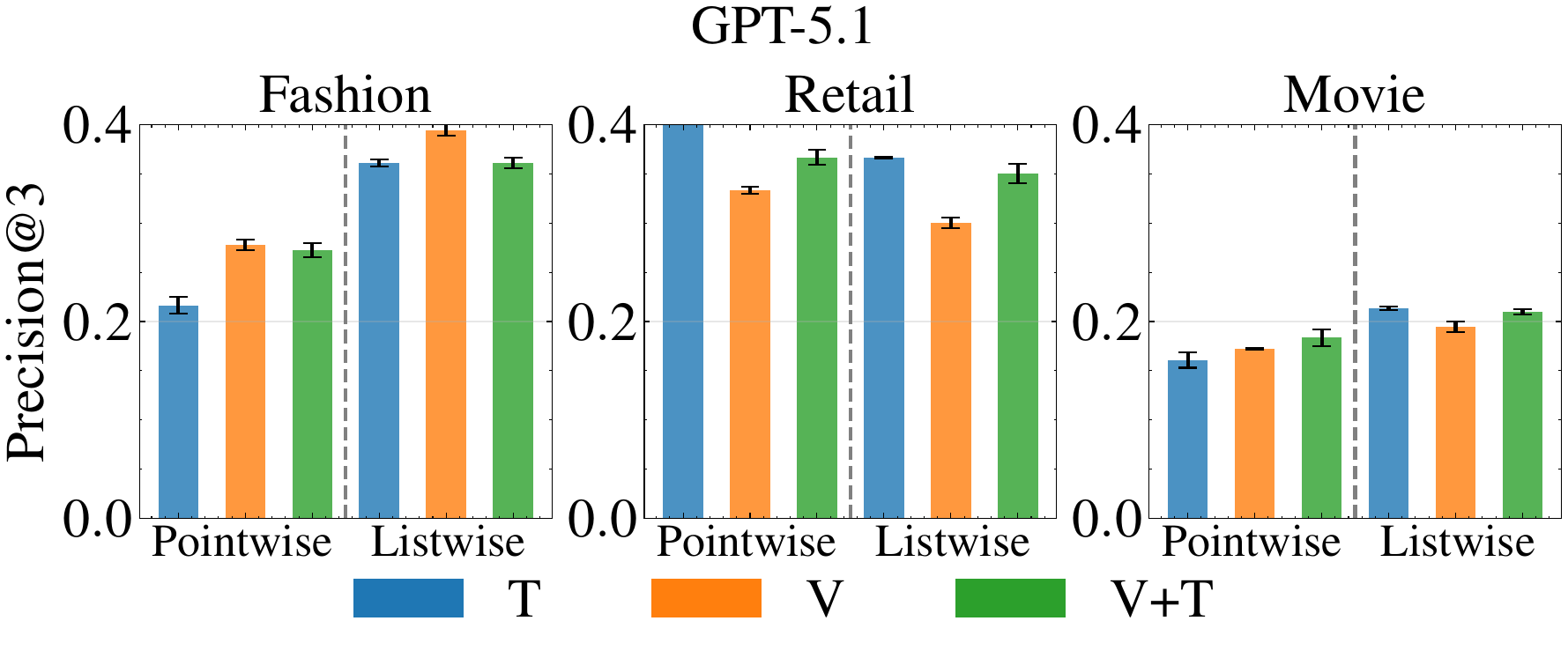}
        \includegraphics[width=\linewidth]{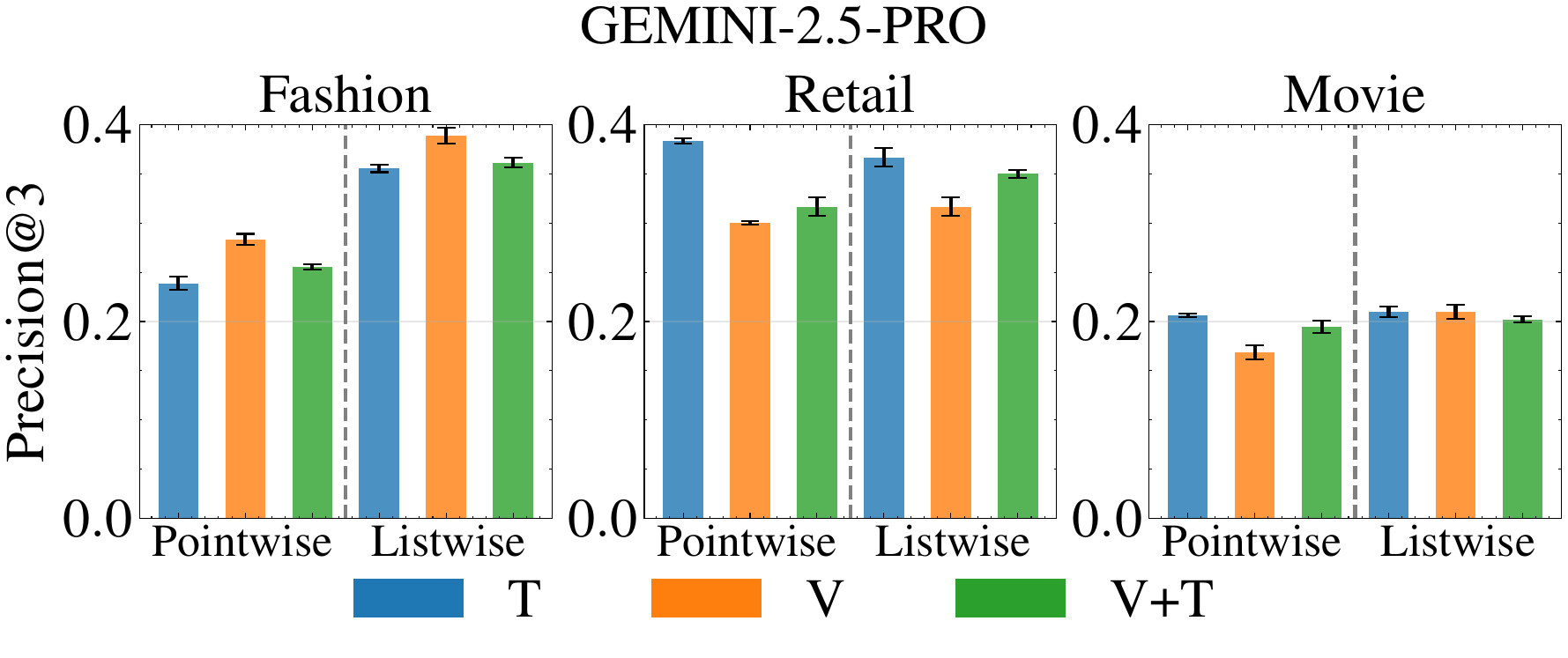}
            \includegraphics[width=\linewidth]{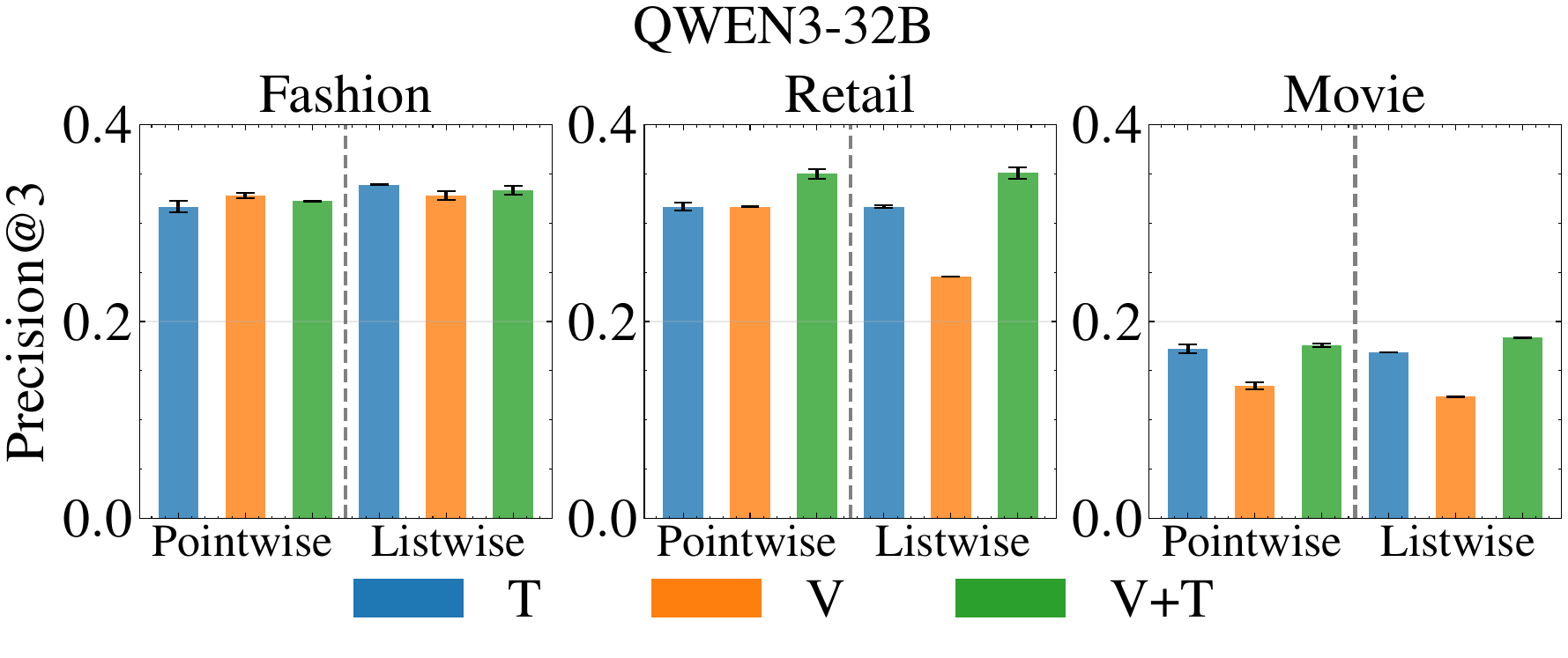}
\caption{Zero-shot VLM as CRS using text(T), visual(V), and combined(V+T) item information across backbones. Visual cues are more informative in Fashion, while textual attributes dominate in Retail.}
    \label{fig:item_reprsentation}
\end{figure}

We further evaluate VLM-based CRS when incorporating items’ visual segmentation. \Autoref{fig:item_reprsentation} reports \texttt{P@3} under LLM-style text-only (T), visual-only (V), and combined visual–textual (V+T) settings.


We observe strong scenario-dependent effects. In the Fashion scenario, visual information of the item substantially improves performance, with visual-only outperforming textual attributes. This suggests that egocentric scenes could provide a discriminative visual signal for items, beyond what their metadata provides. 

In contrast, for the Movie and Retail scenarios, visual information yields limited or even negative gains. This disparity reflects the varying discriminative power of visual cues across scenarios: in retail settings, items on the same shelf are often visually similar and can only be distinguished through non-visual attributes (e.g., function, price, reviews), as illustrated by the car floor mat example in \Autoref{fig:three-plots}.

\textit{Limitation 1:} \emph{Existing methods cannot effectively leverage modalities of item information germane to the scenario. Naively combining visual and textual features (V+T) can be inefficient and may even degrade performance (cf. \Autoref{fig:item_reprsentation}).}


\begin{figure*}
    \centering
    \includegraphics[width=\linewidth]{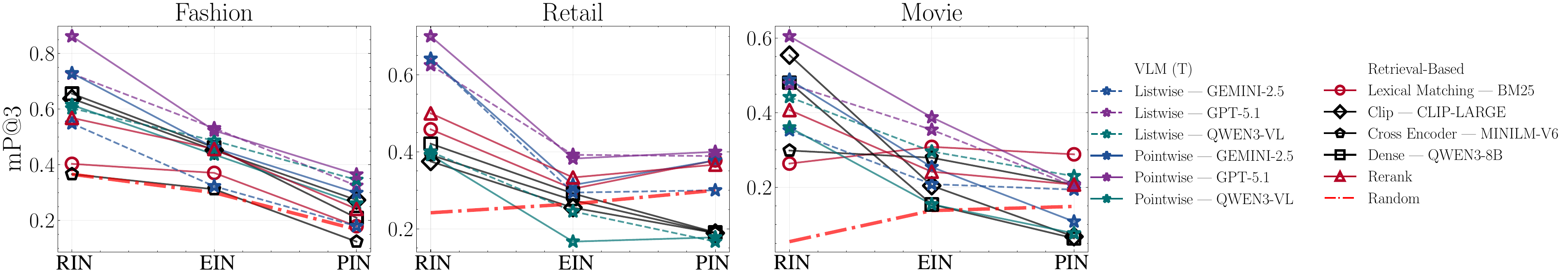}
\caption{Performance of existing immersive label selection methods under three defined criteria across scenarios. While methods achieve strong performance on \textit{RIN}, performance degrades substantially on \textit{EIN} and \textit{PIN}.}
    \label{fig:three_config}
\end{figure*}

\subsection{Immersive Label Selection}

\paragraph{Methods.}
To our knowledge, no existing work is specifically designed to select immersive labels in \imnl{}. We thus adapted existing methods from two representative paradigms (cf. details in \Autoref{app:immserive_label_methods}). 

\begin{figure}
    \centering
\includegraphics[width=\linewidth]{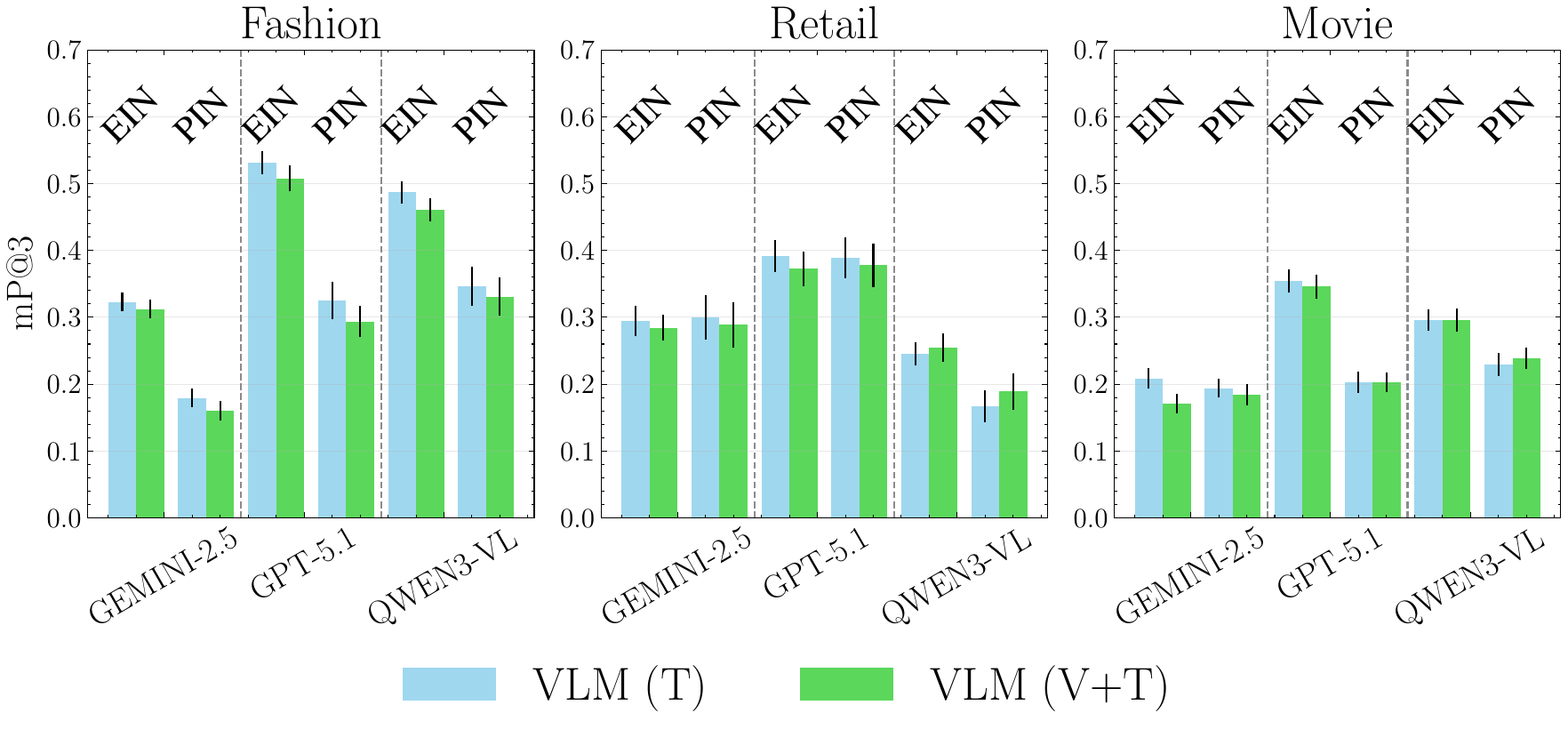}
\caption{Performance of listwise VLM given attributes alone (\textsc{T}) vs with item segmentation (\textsc{V+T}). We observed no noticeable difference by introducing V+T to support latent awareness.}
\label{fig:t_vt}
\end{figure}

\noindent\underline{\textit{Retrieval.}}
Retrieval-based methods treat Label Selection as a relevance scoring problem,
$f(\mathcal{C}_{1:t}, a^{(i)}_k) \!\rightarrow\! \mathbb{R}$, where higher scores indicate stronger alignment between an attribute $a^{(i)}_k$ and $\mathcal{C}_{1:t}$. Instantiations of $f(\cdot)$ include lexical matching, dense retrieval, cross-encoder, re-ranking, or fuse CLIP-based similarity between $\{a^{(i)}_k,\mathcal{V}^{(i)}\}$ with dense similarity between $\{a^{(i)}k,\mathcal{C}_{1:t}\}$ to favor latent but aligned attributes.

\noindent\underline{\textit{Zero-shot VLM}}
Given $\mathcal{C}_{1:t}$ and $\mathcal{A}^{(i)}$, the VLM selects attributes via \emph{pointwise} judgment over each $a^{(i)}_k$ or \emph{listwise} ranking of $\mathcal{A}^{(i)}$. We further assess whether the given $\mathcal{V}^{(i)}$ improves latent awareness. 

We instruct the VLM with the general scenario information and the task definition for either \textit{RIN} or \textit{IN}(cf. \Autoref{sec:item_label}) (evaluated in independent trials). We avoid fine-tuning or explicit guidance on latent or proactive information requirements to assess zero-shot reasoning 
We use the same backbone as for the Item Recommendation task.

\paragraph{Overall Performances in Three Criteria.}
In \Autoref{fig:three_config}, we report the performance of evaluated methods by giving textual attributes only across three criteria. Both VLM-based and several retrieval-based methods achieve high \texttt{mP@3} in \textit{RIN}, which suggests they can reliably identify and address explicit seeker preferences and constraints.

In contrast, performance drops substantially across methods for both \textit{EIN} and \textit{PIN}. The degradation is most pronounced for \textit{PIN}, where several methods perform comparably to a random top-3 baseline (red dashed line). These results indicate that existing methods struggle to align with human expert judgments and the more challenging \emph{latent proactive} level of information needs.

Thus, we hypothesize that failures stem primarily from the two challenges identified in \Autoref{sec:item_label}: (1) limited \emph{latent awareness} and (2) insufficient \emph{proactive disclosure}.
We restrict the evaluation to VLM-based methods (cf. \Autoref{app:pointwise} for pointwise VLM) since retrieval-based methods inherently lack the required reasoning ability.


\paragraph{Latent Awareness.}
First, \Autoref{fig:t_vt} examines whether explicitly providing the item's visual information (T+V) helps systems avoid visually inferable attributes. We observe no performance gains from T+V, suggesting that visual information alone is insufficient and that the VLMs may not intrinsically recognize that the information presented to the seeker should target latent information needs.

\begin{figure}
    \centering
  \includegraphics[width=\linewidth]{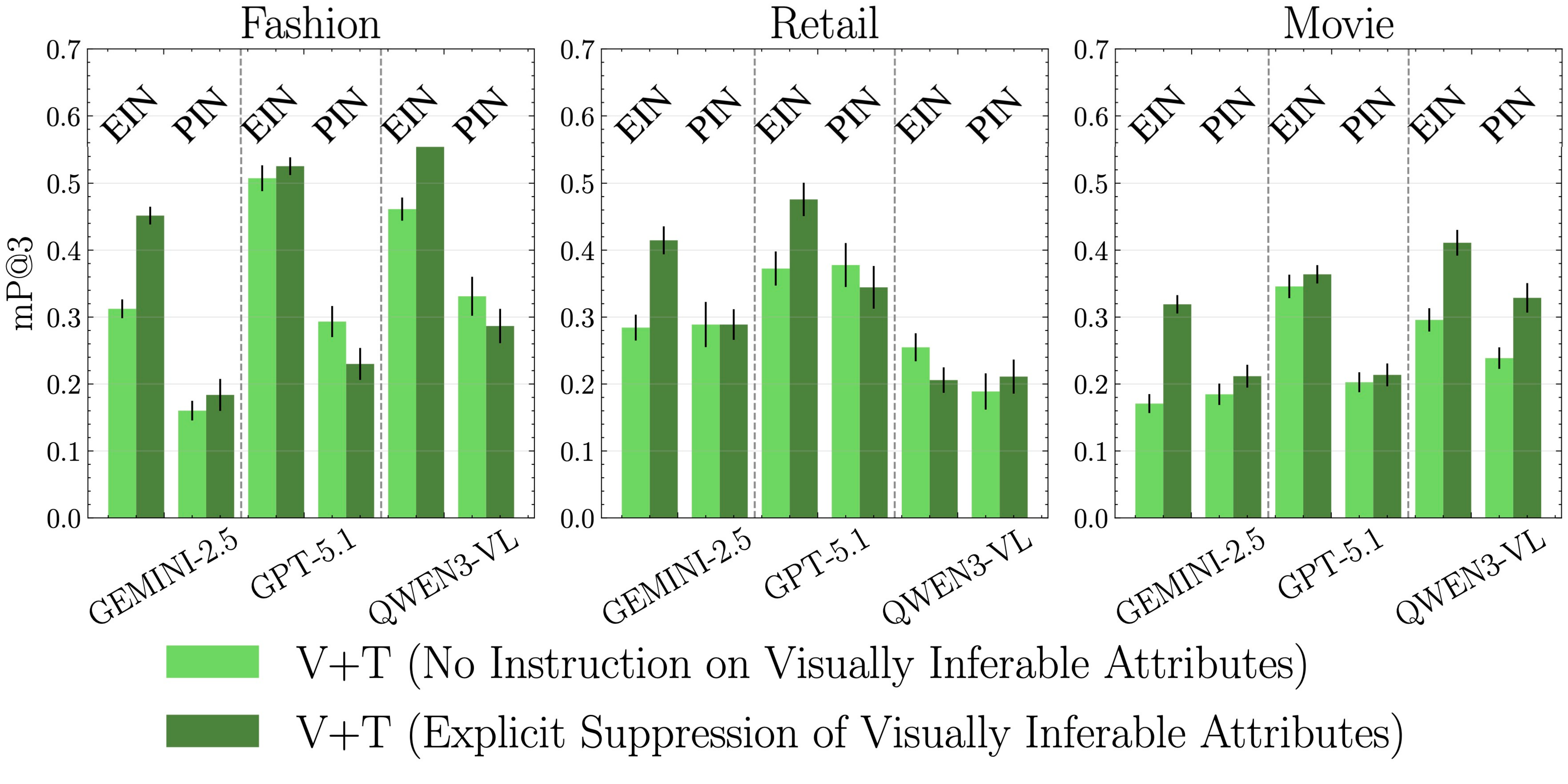}
\caption{VLMs with item visual information (\textsc{V+T}) under instructions to explicitly suppress visually inferable attributes (or not), showing limited latent awareness.}
\label{fig:insturct}
\end{figure}

To validate this hypothesis, we use the few-shot instruction prompt in \Autoref{app:instruct} with the VLM to suppress visually inferable attributes (cf. \Autoref{fig:insturct}). Performance improves in the Movie scenario, where posters often encode attributes (e.g., titles, actors), making systems more prone to selecting visually inferable information over other scenarios.  While few shot instructions can improve latent awareness, this requires scenario-specific few shot design.

\textit{Limitation 2: Existing methods struggle to recognize latent information needs, resulting in redundant presentation of visually inferable information.}

On the contrary, some methods show no performance gains, especially on \textit{PIN}, thus also indicating failures in inferring visually inferable attributes or the presence of confounding factors.

\paragraph{Proactive Disclosure.}
We next examine whether this limitation stems from the incorrect inference of unstated information needs. To isolate this factor, we augment the current set of utterances $\mathcal{C}_{1:t}$ with the subset of utterances in $\mathcal{C}_{t:T}$ annotated with intent tags defined in Sec.~\ref{sec:item_label}. The task reduces to retrieving attributes relevant to explicit requests, rather than inferring unstated information needs.

\begin{figure}
    \centering
\includegraphics[width=\linewidth]{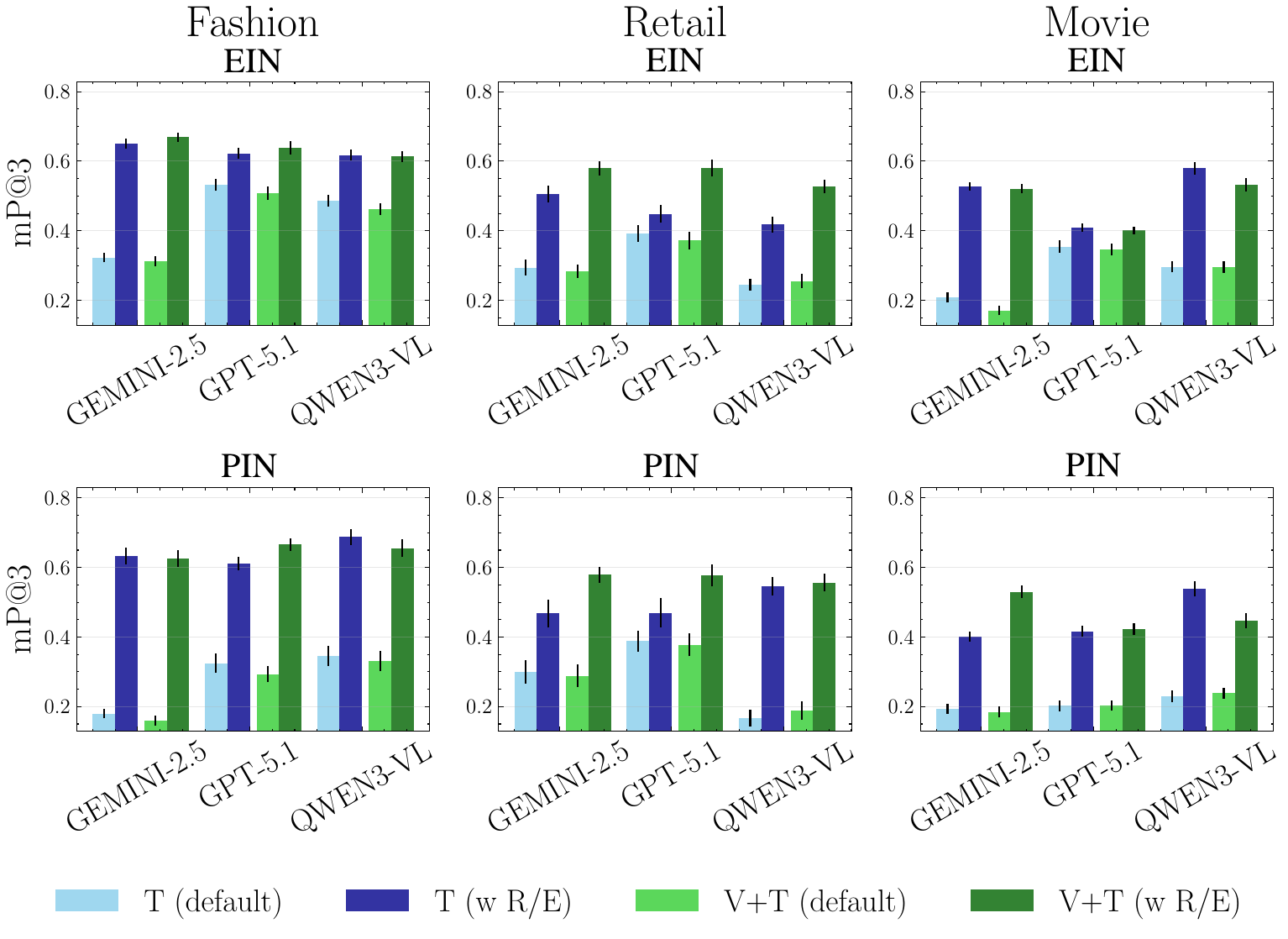}
\caption{Listwise VLM-based methods given utterances tagged as \textit{Implicit Seeker Request} and \textit{Expert Explanation} (\textsc{w R/E}), which simplifies inferring proactive needs into matching explicit requests.}
\label{fig:relax_conversations}
\end{figure}

In \Autoref{fig:relax_conversations}, this setting yields a substantial gain in \texttt{mP@3} compared to the default setting, indicating that a primary source of decline lay in failures to adequately account for unstated information needs. This trend mirrors \textit{RIN}, suggesting systems still struggle with inferring unstated needs, despite reliably retrieving explicitly stated information.

\textit{Limitation 3:}
\emph{Existing methods remain limited in preemptively anticipating seekers’ proactive information needs that are not explicitly stated in the current conversation prefix.} 

\paragraph{False-positive Cases w.r.t. IN-S.} 

We observe that evaluating a composite of false-positive cases across three criteria would obscure limitations in proactive need anticipation, as most methods already perform strongly on retroactive label selection (cf. high \textit{RIN} in \Autoref{fig:three_config}), which would otherwise dominate overall performance. 

Accordingly, we focus our error analysis on \textit{PIN}, the most reasoning-demanding criterion where performance is weakest. We categorize false-positive attributes selected by listwise VLM methods into the following types:
\begin{itemize}
    \item \textit{Visual Inferable (VI):} attributes that can be directly inferred from the visual information.
    \item \textit{Explicit Response (ER):} attributes that respond to explicitly stated seeker questions or recommendation constraints in the conversation prefix.
    \item \textit{Misaligned Proactive (IP):} attributes that are proactive but misaligned with the ground-truth labels for unstated information needs.
\end{itemize}

\begin{figure}
    \centering
    \includegraphics[width=\linewidth]{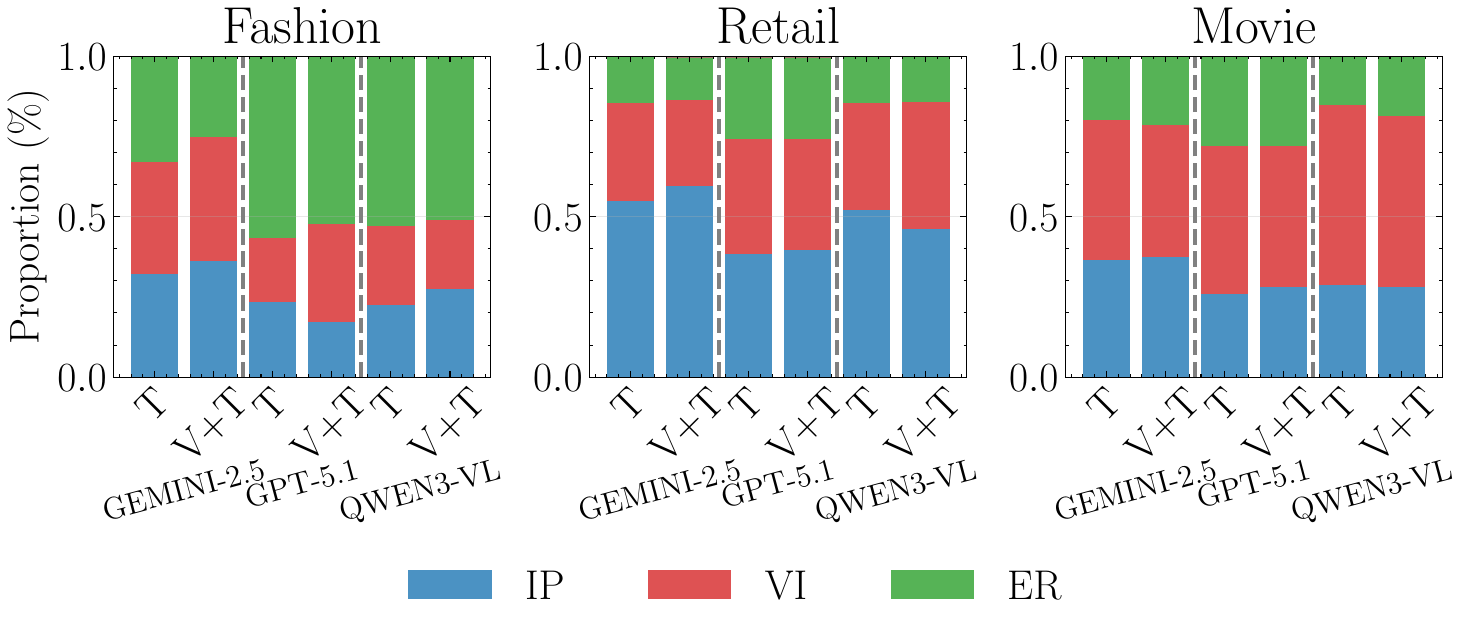}
    \caption{Distribution of attribute false-positive types among the top-3 selections in \textit{PIN} for listwise VLM methods via the default prompt setting.}
    \label{fig:fp_figure}
\end{figure}

\noindent We annotate each false-positive label in the top-3 selections (cf. \Autoref{fig:fp_figure}) with one of VI, ER, or IP. 

\textit{VI} dominates in the Movie scenario, where systems tend to select titles (e.g., \texttt{title: Ready or Not}) or names (e.g., \texttt{director: Tim Miller}) that are often directly inferable from the poster alone. Consequently, few-shot prompting to instruct the system to avoid visually inferable attributes yields measurable performance gains.

As noted in Sec.~\ref{sec:item_rec}, the perceptual clutter constraint limits the number of immersive labels that can be displayed; each selected label must therefore convey high-value, non-visual information. Surfacing visually inferable attributes under this constrained display budget directly reduces label utility, underscoring the importance of latent awareness for effective immersive label selection.

In contrast, the high prevalence of \textit{ER} in Fashion and \textit{IP} in Retail indicates a lack of focus on proactive disclosure. Systems often select attributes (e.g., \texttt{product description: Waterproof protection}) to justify an explicit request (e.g., hiking after rain) stated in the conversation prefix, while failing to anticipate needs in the suffix.


\section{Conclusion}
In this work, we introduce Immersive CRS (\imnl{}) 
that \imnl{} highlights recommended items and presents associated \textit{in-situ} immersive labels within the seeker’s POV to justify explicit intents and address unstated information needs. We propose a novel evaluation framework for immersive labels and benchmark existing methods across multiple  \imnl{} scenarios. Our results reveal that existing methods show limited latent awareness and anticipation of users' latent information needs, often resulting in redundant or uninformative labels that highlight key challenges for future \imnl{} research.

\section*{Acknowledgments}
This work was supported by the Institute of Information \& Communications Technology Planning \& Evaluation (IITP) grant funded by the Korean Government (MSIT) (No. RS-2024-00457882, National AI Research Lab Project).
\section*{Limitations}

First, to support scalable and controlled benchmarking, we adapt existing CRS datasets and synthesize egocentric scenes rather than collecting in-the-wild XR recordings. While this preserves the original conversational signal and yields realistic candidate pools, real-world environments may introduce additional visual and interaction variability.

Second, ground-truth immersive labels are derived from intent-tag annotations over full conversations and adjudicated by an VLM judge given both conversational context and item appearance. Despite following consistent annotation guidelines and restricting labels to atomic attributes, some degree of annotation noise may remain.

Third, our evaluation focuses on a low information-load regime aligned with XR display constraints, considering a limited set of scenarios and representative backbone models. Extending the benchmark to broader scenarios, interface budgets, and future model families is a natural direction for future work.
\bibliography{custom}
\newpage
\appendix
\newpage
\clearpage

\section{Dataset Adaptation}
\label{app:dataset_adaptation}
To enable evaluation under the immersive conversational setting of \imnl{}, we adapt existing CRS datasets using a unified curation pipeline grounded in the formal definitions introduced in \Autoref{sec:crs_prelim}.


\paragraph{Conversation.} 

We retain the original provided conversations $\mathcal{C} = \{(u_t, s_t, \mathcal{I}_t)\}_{t=1}^{T}$ for each CRS  datasets. We filter them to ensure that each preserved conversation contains at least 30 turns and includes at least one utterance for each intent tag defined in \Autoref{par: label_gt} (i.e., \textit{explicit seeker intent}, \textit{implicit user intent}, and \textit{expert explanation}), thereby ensuring that each conversation contains enough meaningful content for our \imnl{} framework.

\paragraph{Candidate Items.}
For each conversation $\mathcal{C}$, we start from the original candidate set $\mathcal{I}^{\text{glb}}$ provided by the CRS dataset. Since datasets such as ReDial, INSPIRED, and CoSRec contain unrealistically large candidate pools under the \imnl{} assumption that items are drawn from the user’s POV view, we apply stratified sampling to obtain $\mathcal{I}^{\text{env}} \subseteq \mathcal{I}^{\text{glb}}$, which contains a realistic number of items for constructing an egocentric scene.
This procedure preserves scenario-level item frequency while ensuring inclusion of the ground-truth item $i_{\mathcal{E}}^\ast \in \mathcal{I}_{\mathcal{E}}$.

\paragraph{Egocentric Scene.} 
For a given conversation $\mathcal{C}$ and a candidate item set $\mathcal{I}^{\text{env}}$, we then construct an egocentric scene $\mathcal{E}$ that depicts how the items in $\mathcal{I}^{\text{env}}$ would appear in an immersive setting (e.g., retail shelves) using \texttt{gemini-2.5-flash-image}. Examples of scenes are illustrated in \Autoref{fig:three-plots}.

\paragraph{External Attributes.}
VOGUE and CoSRec already provide item attributes and product images, either curated from the Amazon website or linked to the Amazon Product Search Corpus. However, ReDial and INSPIRED do not include high-quality movie posters or detailed movie attributes. Therefore, we augment these datasets by crawling additional movie metadata, reviews, and posters using the TMDB API\footnote{\url{https://developer.themoviedb.org}}. Then, for each item in each dataset, we apply the decomposition rules described in \Autoref{sec:item_label} to obtain a semantically minimal but self-contained set of attribute snippets.

\paragraph{Item Recommendations.}  
Given a conversation prefix $\mathcal{C}_{1:t}$, the ground truth item set 
$\mathcal{I}_{t}^{\text{env*}} \subseteq \mathcal{I}^{\text{env}}$ is directly adapted from the original dataset.

\subsection{Atomic Attribute Decomposition for Immersive Label Selection.}
For each recommended item $i$, we construct a candidate attribute set $\mathcal{A}^{(i)}$ from the item metadata provided in the adapted CRS datasets.
Each item is associated with a collection of \emph{composite attributes} represented as key--value pairs (e.g., \texttt{material: leather}, \texttt{fit: slim fit}, \texttt{brand: Nike}), which serve as the primary source of descriptive information.

We first decompose each composite attribute into a set of sentence-level attribute descriptions by concatenating the key and value into a natural-language sentence (e.g., \emph{``Material: leather''}).
To further obtain \emph{atomic attributes}, we optionally apply sentence-level splitting using a lightweight NLP pipeline (spaCy) in conjunction with Gemini~2.5~Flash, which decomposes compound descriptions into minimal, semantically independent units when applicable.
Each resulting atomic attribute is augmented with its originating attribute key to preserve semantic provenance.

\vspace{0.5em}
\noindent\textbf{Example 1: Children's Book (\textit{Dinosnores})}

From the original metadata fields (e.g., \texttt{features}, \texttt{description}), we derive the following atomic attributes:
\begin{itemize}[leftmargin=1.5em]
    \item \texttt{title}: \emph{Dinosnores}
    \item \texttt{genre}: Children's book
    \item \texttt{theme}: Whimsical bedtime routine
    \item \texttt{style}: Playful verse
    \item \texttt{illustration}: Vivid and colorful illustrations
\end{itemize}
Each attribute expresses a single interpretable property and does not overlap with others (e.g., narrative theme vs.\ visual style).

\vspace{0.5em}
\noindent\textbf{Example 2: Contemporary Fiction (\textit{When You Read This})}

For a novel with rich narrative descriptions, we decompose high-level summaries into the following atomic attributes:
\begin{itemize}[leftmargin=1.5em]
    \item \texttt{genre}: Contemporary literary fiction
    \item \texttt{format}: Epistolary narrative
    \item \texttt{theme}: Grief and healing
    \item \texttt{tone}: Humorous yet poignant
    \item \texttt{focus}: Interpersonal relationships
\end{itemize}
Notably, composite descriptions such as ``humorous exploration of grief and love'' are decomposed into separate attributes for tone and theme.

\vspace{0.5em}
\noindent\textbf{Example 3: Nonfiction / Instructional Book (\textit{The Mandolin Project})}

Technical and instructional content is similarly decomposed:
\begin{itemize}[leftmargin=1.5em]
    \item \texttt{topic}: Mandolin construction
    \item \texttt{difficulty}: Suitable for first-time builders
    \item \texttt{content}: Step-by-step instructions
    \item \texttt{supplement}: Full-size plans and diagrams
    \item \texttt{style}: Conversational instructional tone
\end{itemize}

\vspace{0.5em}
\noindent

The final candidate set $\mathcal{A}^{(i)}$ thus consists of atomic textual attributes that are concise, non-overlapping, and uniformly formatted across items.
All evaluated methods operate on the same constructed attribute sets, and no further decomposition or merging is performed during inference to ensure fair comparison.

\subsection{Ground Truth Annotation for Immersive Label Selection.} \label{app:ground_truth}

We follow \Autoref{sec:item_label} to construct the ground truth immersive label sets 
$\mathcal{A}_{t}^{(i),\text{crit}*}$ for each evaluation criterion and each item 
$i \in \mathcal{I}_{t}^{\text{env}*}$ given an observed conversation prefix $\mathcal{C}_{1:t}$.  

For each atomic attribute (snippet), we formulate ground truth label relevance with respect to the information that would be available to an oracle recommender at decision point $t$: (i) the observable prefix $\mathcal{C}_{1:t}$ with explicit seeker requests for \textit{RIN}, (ii) the full conversation $\mathcal{C}_{1:T}$ with expert explanations extracted via intent tags from the recommender for \textit{EIN}, or (iii) the full conversation $\mathcal{C}_{1:T}$ with seeker requests extracted via intent tags (e.g., post-recommendation follow-ups) for \textit{PIN}. Conditioned on the relevant evidence slice and item-specific context, we prompt \textsc{GPT-5.1} to judge whether the atomic attribute directly addresses at least one utterance in the corresponding intent set (seeker requests, expert explanations, or seeker requests), and to provide a short rationale. Full prompts are provided below. For cost and time considerations, we batch attribute relevance labeling into groups of 10, for the same item only.

\paragraph{\textit{\textit{RIN}} (Retroactive Information Need).}\mbox{}\\
\textbf{System Prompt:}
\begin{lstlisting}
You judge whether each individual snippet explicitly and directly contains information that matches the seeker's pre-recommendation request.

Inputs:
- A partial conversation transcript assistant seeker interaction.
- A list of pre-recommendation seeker queries aligned to the conversation.
- A list of item snippets. Each snippet is independent.

Non-visual rule: information obvious from product visuals (e.g., color, visible material, logos, number of pockets, style, shape, or pattern) must be marked non-relevant.

Decision process (per snippet):
1) Summarize or quote the seeker request.
2) Decide whether the snippet directly answers that request.

Output JSON array only:
[{ "id": ..., "relevance": 0|1, "reason": "..." }].
\end{lstlisting}

\textbf{User Input:}
\begin{lstlisting}
conversation: <CONVERSATION TRANSCRIPT - UP TO FIRST RECOMMENDATION TURN>
explicit_seeker_requests: <EXPLICIT SEEKER REQUESTS>
item_snippets: <ITEM AND SNIPPET INFORMATION>
\end{lstlisting}

\paragraph{\textit{EIN} (Expert Inferred Information Need).}\mbox{}\\
\textbf{System Prompt:}
\begin{lstlisting}
You judge whether each individual snippet explicitly and directly contains information that supports the recommender's explanations for recommending an item.

Inputs:
- A partial conversation transcript (assistant - seeker interaction).
- A list of pre-recommendation seeker queries aligned to the conversation.
- A list of item snippets. Each snippet is independent.

Non-visual rule: information obvious from product visuals must be marked non-relevant.

Decision process (per snippet):
1) Summarize or quote the assistant explanations.
2) Decide whether the snippet directly supports those explanations.

Output JSON array only:
[{ "id": ..., "relevance": 0|1, "reason": "..." }].
\end{lstlisting}

\textbf{User Input:}
\begin{lstlisting}
conversation: <CONVERSATION TRANSCRIPT>
recommender_explanations: <RECOMMENDER EXPLANATIONS>
item_snippets: <ITEM AND SNIPPET INFORMATION>
\end{lstlisting}

\paragraph{\textit{PIN} (Information Need -- Seeker).}\mbox{}\\
\textbf{System Prompt:}
\begin{lstlisting}
You judge whether each individual snippet explicitly and directly contains information that answers the seeker's questions.

Inputs:
- A partial conversation transcript (assistant - seeker interaction).
- A list of pre-recommendation seeker queries aligned to the conversation.
- A list of item snippets. Each snippet is independent.

Non-visual rule: information obvious from product visuals must be marked non-relevant.

Decision process (per snippet):
1) Summarize or quote the seeker questions.
2) Decide whether the snippet directly answers those questions.

Output JSON array only:
[{ "id": ..., "relevance": 0|1, "reason": "..." }].
\end{lstlisting}

\textbf{User Input:}
\begin{lstlisting}
conversation: <CONVERSATION TRANSCRIPT>
seeker_questions: <SEEKER QUESTIONS>
item_snippets: <ITEM AND SNIPPET INFORMATION>
\end{lstlisting}

\section{Description of Method Adaptation in the Immersive Label Selection}\label{app:immserive_label_methods}

This section provides implementation-level details for all evaluated immersive label selection methods.
\subsection{Retrieval-based Label Selection}
Retrieval-based methods formulate immersive label selection as a relevance scoring problem,
$f(\mathcal{C}_{1:t}, a^{(i)}_k) \rightarrow \mathbb{R}$,
where higher scores indicate stronger alignment between the conversation prefix $\mathcal{C}_{1:t}$ and an attribute $a^{(i)}_k$ of item $i$.

\paragraph{Lexical Retrieval.}
Lexical retrieval uses BM25 scoring between $\mathcal{C}_{1:t}$ and each $a^{(i)}_k \in \mathcal{A}^{(i)}$. Stopwords are retained to preserve conversational structure, and no query expansion or heuristic filtering is applied. Attributes are ranked directly by BM25 score.

\paragraph{Dense Retrieval.}
Dense retrieval encodes the conversation prefix $\mathcal{C}_{1:t}$ and each candidate attribute $a^{(i)}_k$ into a shared dense vector space using a bi-encoder architecture.
Both conversation queries and attribute texts are embedded using the same embedding backend with a fixed dimensionality (default: 1536).
Embeddings are L2-normalized at encoding time, allowing relevance to be computed via inner product, which is equivalent to cosine similarity.

At scoring time, the query embedding is cast to \texttt{float32} and exhaustively scored against all candidate attribute embeddings using a matrix--vector dot product.
Given the moderate size of $\mathcal{A}^{(i)}$, no approximate nearest-neighbor search or pruning is applied.
The resulting relevance scores are sorted in descending order, with ties broken deterministically by attribute identifier, and the top-$k$ attributes are returned.

\paragraph{Cross-Encoder.}
Cross-encoder models jointly encode each conversation--attribute pair $(\mathcal{C}_{1:t}, a^{(i)}_k)$ and directly output a scalar relevance score.
Unlike bi-encoder retrieval, the cross-encoder processes the concatenated input in a single forward pass, enabling full cross-attention between the conversation context and the attribute text.

In our implementation, we use a MiniLM-v6 cross-encoder backbone.
The conversation prefix $\mathcal{C}_{1:t}$ and attribute $a^{(i)}_k$ are serialized into a paired input sequence following the MiniLM-v6 input format.
Each pair is evaluated independently, producing a real-valued relevance score without embedding reuse across attributes.

All candidate attributes $a^{(i)}_k \in \mathcal{A}^{(i)}$ are exhaustively scored, as the candidate set size is sufficiently small to permit full enumeration.
Inference is performed in fixed-size batches with the model in evaluation mode and without gradient computation.
No task-specific fine-tuning is applied; the MiniLM-v6 cross-encoder is used as a zero-shot relevance scorer.
Final rankings are obtained by sorting relevance scores in descending order, with deterministic tie-breaking.

\paragraph{Rerank.}
In this two-stage retrieval--reranking approach, we first apply BM25-based lexical retrieval to obtain an initial candidate subset of attributes.
For each conversation prefix $\mathcal{C}_{1:t}$, all candidate attributes $a^{(i)}_k \in \mathcal{A}^{(i)}$ are scored using BM25 with standard hyperparameters $k_1 = 1.5$ and $b = 0.75$.
The top-$K$ attributes are selected as candidates for re-ranking.

The selected candidates are then re-ranked using the Cohere Rerank v3.5 model.
The reranker jointly evaluates each $(\mathcal{C}_{1:t}, a^{(i)}_k)$ pair and produces a scalar relevance score that captures fine-grained semantic interactions beyond lexical overlap.
Final rankings are obtained by sorting reranker scores in descending order, with deterministic tie-breaking.

The same BM25 configuration, reranker model, and inference protocol are applied consistently across all scenarios.
No task-specific fine-tuning or scenario adaptation is performed.

\paragraph{CLIP-based Visual--Textual Similarity and Fusion.}
To incorporate item visual information into immersive label selection, we use the CLIP model with a ViT-B/32 backbone to compute visual--textual similarity between candidate attributes and item images.
Given an item image $\mathcal{V}^{(i)}$ and a candidate attribute $a^{(i)}_k$, CLIP independently encodes the image and text into a shared embedding space.
All embeddings are L2-normalized, and visual--textual similarity is computed via cosine similarity.

In multimodal retrieval settings, the CLIP-based similarity score is combined with textual similarity between the conversation prefix and the attribute,
$s_{\text{text}}(\mathcal{C}_{1:t}, a^{(i)}_k)$, computed using dense retrieval.
The final relevance score is obtained via fixed linear fusion:
$
s(a^{(i)}_k) \;=\; \lambda \, s_{\text{text}}(\mathcal{C}_{1:t}, a^{(i)}_k)
\;-\; (1-\lambda)\, s_{\text{CLIP}}(a^{(i)}_k, \mathcal{V}^{(i)}),
$
where $\lambda$ is a fixed weight shared across all scenarios and experiments.

Importantly, this fusion is designed to emphasize semantic alignment between the conversation and candidate attributes while discouraging the selection of attributes that are strongly grounded in visual information alone.
Since dense retrieval prioritizes attributes that match conversational intent, and CLIP assigns higher similarity to visually inferable attributes, attributes that are semantically relevant to the conversation but weakly supported by visual grounding are preferentially promoted.
Conversely, attributes that are visually salient but weakly aligned with the conversation are down-weighted.

As a result, the fusion mechanism implicitly favors attributes whose relevance cannot be explained solely by visual information, encouraging the selection of non-obvious, latent information that better satisfies the user’s underlying decision-making needs.
We emphasize that this behavior arises from the complementary roles of the two similarity terms and does not rely on any learned objective, fine-tuning, or scenario-specific adaptation.
All fusion weights are fixed \emph{a priori}, and CLIP is used strictly in inference mode.

\subsection{Zero-shot VLM (Pointwise).}
In the pointwise setting, immersive label selection is formulated as an independent relevance judgment problem.
For each candidate attribute $a^{(i)}_k \in \mathcal{A}^{(i)}$, the VLM is queried separately given the conversation prefix $\mathcal{C}_{1:t}$ and a single attribute description.
The model is instructed to assess whether the attribute is relevant under the specified objective (either \textit{RIN} or \textit{IN}) and to output a scalar relevance score or binary decision.

Each attribute is evaluated in an independent forward pass, and no information about other candidate attributes is provided to the model.
This design explicitly prevents cross-attribute comparison or competition, isolating the model’s ability to judge relevance based solely on local evidence.
Outputs are parsed deterministically and used to rank attributes by descending score (or by positive judgments when binary outputs are used).

For VLMs, the item image $\mathcal{V}^{(i)}$ is provided alongside the textual prompt. Images are resized and normalized according to the backbone defaults. No region-level annotations or visual prompts are provided, ensuring a fair comparison with text-only VLMs.

To ensure reproducibility, we use greedy decoding with temperature set to zero and fixed prompt templates across all scenarios.
No few-shot examples, fine-tuning, or explicit guidance on latent or proactive information needs are provided, allowing us to evaluate the zero-shot reasoning capability of VLMs in the pointwise regime.

\paragraph{General Prompt Design.}
The prompt explicitly conditions the model on three elements:
(i) the conversation prefix $\mathcal{C}_{1:t}$ between the seeker and the system,
(ii) a set of candidate attribute snippets describing the recommended item, and
(iii) optionally, a visual segment of the item's visual information.
To reflect the immersive setting, the prompt instructs the model \emph{not} to surface information that is visually obvious (e.g., color, style, or product name), as such information is directly observable by the user through the XR interface.

The core decision criterion is parameterized to support two evaluation objectives.
When evaluating \textit{RIN}, the model is asked to assess whether an attribute explicitly addresses the seeker’s stated requests or preferences.
When evaluating \textit{IN}, the model instead judges whether an attribute provides additional information that proactively supports decision-making, even if it was not explicitly requested.
\textit{These two objectives are evaluated in separate runs using the same prompt structure.}

\paragraph{Scoring Scheme.}
Each attribute snippet is assigned an integer relevance score on a four-point ordinal scale:
0 (irrelevant), 1 (weak evidence), 2 (partial relevance), and 3 (strong relevance).
This coarse-grained scale encourages consistent ranking behavior while remaining interpretable.
All candidate snippets are scored independently, but presented together to allow the model to calibrate relevance across the set.

\paragraph{Output Format and Constraints.}
For reproducibility and deterministic parsing, the model is required to return \emph{only} a JSON array.
Each entry contains the snippet identifier, its relevance score, and a brief rationale justifying the score.
All provided snippet IDs must appear exactly once in the output, and no additional text is permitted outside the JSON structure.
This strict output constraint enables automated evaluation and prevents post-hoc filtering..

\paragraph{Prompt Template}
The same prompt structure is applied across all scenarios and models, where \textit{RIN} (explicit satisfaction) and \textit{IN} (implicit/proactive need) are instantiated in separate runs by switching the \texttt{Evaluation Objective} block.

\begin{lstlisting}[language={}, basicstyle=\ttfamily\scriptsize, breaklines=true]
You are an assistant for an immersive conversational recommendation system.
The user is wearing AR glasses and is physically present in a store.
The recommended item is already visible in the user's field of view.

Context:
You are given:
1. A conversation between the system and the user.
2. A set of attributes describing the recommended item.
3. (Optional) A visual segment of the item's visual information.

Your task:
For EACH attribute, decide whether it should be shown as an immersive
textual label to support the user's decision-making in this physical setting.

Evaluation Objective:
- If the objective is \textit{\textit{RIN}} (Retroactive Information Need):
  Determine whether the snippet directly answers or addresses information
  that the user has explicitly requested, stated, or confirmed in the
  conversation.

- If the objective is IN (Implicit Information Need):
  Determine whether the snippet provides additional, non-obvious information
  that would proactively help the user make a better decision, which
  was not explicitly requested in the conversation.

Conversation:
{CONVERSATION_PREFIX}

Scoring Scale:
0 = Clearly irrelevant or does not satisfy the evaluation objective
1 = Weak relevance or marginally related information
2 = Moderately relevant but not among the most helpful attributes
3 = Highly relevant and clearly satisfies the evaluation objective

Instructions:
You will be given {N} attribute snippets.
For EACH snippet, assign a relevance score from 0 to 3.

Output Format:
Return ONLY a JSON array in the following exact format:
[
  {"id": "1", "relevance": 3, "rationale": "brief justification"},
  {"id": "2", "relevance": 1, "rationale": "brief justification"},
  ...
]

Output Requirements:
- Include ALL snippet IDs exactly once.
- Use the exact IDs provided below.
- Provide a concise rationale grounded in the evaluation objective.
- Do NOT include explanations outside the JSON array.

SNIPPETS:
[1] SNIPPET_ID: {ID_1}
SNIPPET_TEXT: {TEXT_1}
---
[2] SNIPPET_ID: {ID_2}
SNIPPET_TEXT: {TEXT_2}
---
...
\end{lstlisting}

\paragraph{Implementation-Specific Details.}
In implementation, the prompt is dynamically instantiated with:
(i) the full conversation prefix serialized as plain text,
(ii) the number of candidate snippets,
(iii) the exact snippet IDs and snippet texts, and
(iv) a flag indicating whether item visual information is included.
The same prompt template is used across all scenarios and models.
No few-shot examples, chain-of-thought instructions, or task-specific fine-tuning are applied.
All VLM inference is performed with greedy decoding (temperature set to zero) to minimize stochastic variation.

\subsection{Zero-shot VLM (Listwise).}
In the listwise setting, immersive label selection is formulated as a joint ranking problem over the full candidate attribute set $\mathcal{A}^{(i)}$.
The entire set of attribute snippets associated with item $i$ is provided to the VLM in a single prompt, together with the conversation prefix $\mathcal{C}_{1:t}$.
The model is instructed to compare attributes against each other and either rank them by relevance or select the top-$k$ attributes that best satisfy the specified evaluation objective (\textit{RIN} or \textit{IN}).

Unlike the pointwise formulation, the listwise setting allows the model to reason over relative importance and competition among candidate attributes, enabling more coherent selection under a constrained display budget.
However, the model is not provided with any explicit supervision on latent or proactive reasoning beyond the task description.

Model outputs are required to follow a strict, machine-parseable format (e.g., an ordered list of attribute identifiers).
Rankings are parsed deterministically, and no post-processing heuristics, score calibration, or reranking are applied.
All evaluations use greedy decoding with fixed prompt templates across scenarios to ensure reproducibility and isolate zero-shot listwise reasoning behavior.

\paragraph{Prompt Template.}
The same prompt structure is applied across all scenarios and models, where \textit{RIN} (explicit satisfaction) and \textit{IN} (implicit/proactive need) are instantiated in separate runs by switching the \texttt{Evaluation Objective} block.

\begin{lstlisting}[language={}, basicstyle=\ttfamily\scriptsize, breaklines=true]
You are an assistant for an immersive conversational recommendation system.
The user is wearing AR glasses and is physically present in a store.
The recommended item is already visible in the user's field of view.

Context:
You are given:
1) A conversation between the system and the user.
2) A set of attribute snippets describing the recommended item.
3) (Optional) A visual segment of the item visual information.

Your task:
Rank the attribute snippets by how suitable they are to display as immersive
textual labels to support the user's decision-making in this physical setting.

Evaluation Objective:
- If the objective is \textit{RIN} (Retroactive Information Need):
  Prefer snippets that directly answer or address the user's explicitly stated
  requests, questions, or preferences in the conversation.

- If the objective is PIN (Proactive Information Need):
  Prefer snippets that provide additional, non-obvious information that would
  proactively help the user make a better decision which
  was not explicitly requested in the conversation.

Conversation:
{CONVERSATION_PREFIX}

Ranking Instruction:
You will be given N attribute snippets. Rank them from most suitable to least
suitable under the evaluation objective and constraints above.

Return ONLY the TOP 5 snippets, ranked 1 (most suitable) to 5 (fifth).

Output Format:
Return ONLY a JSON array with this exact format:
[
  {"id": "<INFORMATION_ID>", "rank": 1, "rationale": "brief justification"},
  {"id": "<INFORMATION_ID>", "rank": 2, "rationale": "brief justification"},
  {"id": "<INFORMATION_ID>", "rank": 3, "rationale": "brief justification"},
  {"id": "<INFORMATION_ID>", "rank": 4, "rationale": "brief justification"},
  {"id": "<INFORMATION_ID>", "rank": 5, "rationale": "brief justification"}
]

Output Requirements:
- Include exactly 5 entries.
- Use the exact INFORMATION_ID values provided below.
- Rationale must be grounded in the evaluation objective and constraints.
- Do NOT include any text outside the JSON array.

SNIPPETS:
[1] INFORMATION_ID: {ID_1}
INFORMATION_TEXT: {TEXT_1}
---
[2] INFORMATION_ID: {ID_2}
INFORMATION_TEXT: {TEXT_2}
---
...
\end{lstlisting}

\section{Few Shot Instruction Prompt}\label{app:instruct}
The following prompts instantiate scenario-specific instructions to discourage the selection of visually inferable attributes.
They are applied only in controlled few-shot settings to analyze the impact of explicit constraints on VLMs, and are not used in the default zero-shot evaluation. They are appended to original prompt of the specific VLM based methods selected.

\paragraph{Fashion.} The prompt used for experiments in Fashion dataset is

\begin{lstlisting}[language={}, basicstyle=\ttfamily\scriptsize, breaklines=true]
Important Constraint:
Do NOT select attributes that can be directly inferred from the item's visual information.

Examples of visually inferable attributes in Fashion:
1. "The jacket is black and has a slim silhouette."
2. "The dress features a floral pattern with long sleeves."
3. "The shoes have a low-profile design with visible branding."

Assume the user can already see color, shape, texture, pattern, and overall style.

Focus instead on attributes that provide non-visual information
(e.g., material properties, comfort, durability, care, or usage conditions).
\end{lstlisting}

\begin{table*}
\renewcommand{\arraystretch}{1.2}
\scriptsize
\centering
\resizebox{\linewidth}{!}{
    \begin{tabular}{lll|lll|lll|lll}
\toprule
 &  & scenario & \multicolumn{3}{c}{\textbf{Fashion}} & \multicolumn{3}{c}{\textbf{Retail}} & \multicolumn{3}{c}{\textbf{Movie}} \\
 &  & metric & P@1 & P@2 & P@3 & P@1 & P@2 & P@3 & P@1 & P@2 & P@3 \\
method & form & model &  &  &  &  &  &  &  &  &  \\
\hline
BM25 &  &  & 0.150±0.005 & 0.108±0.011 & 0.100±0.009 & 0.100±0.008 & 0.225±0.003 & 0.183±0.003 & 0.101±0.002 & 0.135±0.011 & 0.135±0.008 \\
\hline
Dense &  & QWEN3-8B & 0.183±0.009 & 0.192±0.001 & 0.200±0.012 & 0.300±0.010 & 0.350±0.003 & 0.267±0.003 & 0.112±0.003 & 0.129±0.004 & 0.124±0.007 \\
\hline
Rerank &  &  & 0.050±0.008 & 0.042±0.003 & 0.078±0.002 & 0.150±0.011 & 0.175±0.012 & 0.183±0.010 & 0.112±0.004 & 0.124±0.002 & 0.131±0.009 \\
\hline
\multirow[t]{6}{*}{LLM(T)} & \multirow[t]{3}{*}{pointwise} & GEMINI-2.5P & 0.250±0.006 & 0.275±0.002 & 0.239±0.006 & 0.500±0.001 & \textbf{0.475±0.011} & 0.383±0.004 & 0.270±0.008 & 0.230±0.004 & 0.206±0.007 \\
 &  & GPT-5.1 & 0.193±0.007 & 0.211±0.003 & 0.216±0.012 & 0.550±0.010 & \textbf{0.475±0.011} & \textbf{0.417±0.011} & 0.157±0.008 & 0.180±0.011 & 0.161±0.002 \\
 &  & QWEN3-VL & 0.333±0.003 & 0.325±0.001 & 0.317±0.005 & 0.450±0.005 & 0.375±0.004 & 0.317±0.010 & 0.135±0.005 & 0.174±0.004 & 0.172±0.007 \\
\cline{2-12}
 & \multirow[t]{3}{*}{listwise} & GEMINI-2.5P & 0.450±0.003 & 0.383±0.010 & 0.356±0.002 & 0.400±0.012 & 0.400±0.009 & 0.367±0.003 & 0.202±0.001 & 0.225±0.010 & 0.210±0.009 \\
 &  & GPT-5.1 & 0.450±0.009 & 0.400±0.009 & 0.361±0.002 & 0.400±0.005 & 0.375±0.002 & 0.367±0.010 & 0.247±0.008 & \textbf{0.270±0.005} & \textbf{0.213±0.002} \\
 &  & QWEN3-VL & 0.417±0.004 & 0.400±0.005 & 0.339±0.009 & 0.400±0.008 & 0.350±0.011 & 0.317±0.006 & 0.112±0.002 & 0.163±0.009 & 0.169±0.009 \\
\hline
\multirow[t]{6}{*}{VLM(V+T)} & \multirow[t]{3}{*}{pointwise} & GEMINI-2.5P & 0.317±0.007 & 0.308±0.009 & 0.256±0.006 & 0.500±0.007 & 0.375±0.006 & 0.317±0.001 & 0.213±0.002 & 0.213±0.001 & 0.195±0.008 \\
 &  & GPT-5.1 & 0.283±0.004 & 0.300±0.007 & 0.272±0.011 & \textbf{0.650±0.004} & 0.425±0.006 & 0.367±0.009 & 0.169±0.004 & 0.213±0.002 & 0.184±0.004 \\
 &  & QWEN3-VL & 0.383±0.003 & 0.367±0.011 & 0.322±0.010 & 0.400±0.008 & 0.375±0.011 & 0.350±0.010 & 0.157±0.003 & 0.191±0.011 & 0.176±0.007 \\
\cline{2-12}
 & \multirow[t]{3}{*}{listwise} & GEMINI-2.5P & \textbf{0.533±0.010} & \textbf{0.475±0.011} & 0.361±0.004 & 0.450±0.002 & 0.375±0.004 & 0.350±0.006 & 0.202±0.010 & 0.225±0.010 & 0.202±0.001 \\
 &  & GPT-5.1 & 0.417±0.007 & 0.417±0.006 & 0.361±0.003 & 0.550±0.002 & 0.425±0.005 & 0.350±0.011 & \textbf{0.337±0.005} & 0.253±0.007 & 0.210±0.009 \\
 &  & QWEN3-VL & 0.417±0.005 & 0.383±0.012 & 0.333±0.012 & 0.316±0.004 & 0.368±0.006 & 0.351±0.004 & 0.090±0.004 & 0.163±0.001 & 0.184±0.008 \\
\hline
\multirow[t]{6}{*}{VLM(V)} & \multirow[t]{3}{*}{pointwise} & GEMINI-2.5P & 0.200±0.007 & 0.267±0.002 & 0.283±0.004 & 0.450±0.011 & 0.350±0.004 & 0.300±0.003 & 0.191±0.006 & 0.191±0.012 & 0.169±0.004 \\
 &  & GPT-5.1 & 0.200±0.008 & 0.275±0.009 & 0.278±0.004 & 0.550±0.009 & 0.400±0.005 & 0.333±0.008 & 0.191±0.008 & 0.202±0.007 & 0.172±0.002 \\
 &  & QWEN3-VL & 0.317±0.010 & 0.333±0.005 & 0.328±0.003 & 0.500±0.001 & 0.375±0.007 & 0.317±0.008 & 0.124±0.001 & 0.124±0.007 & 0.135±0.003 \\
\cline{2-12}
 & \multirow[t]{3}{*}{listwise} & GEMINI-2.5P & 0.450±0.008 & 0.467±0.003 & 0.389±0.009 & 0.200±0.005 & 0.250±0.011 & 0.317±0.003 & 0.258±0.005 & 0.242±0.002 & 0.210±0.011 \\
 &  & GPT-5.1 & 0.517±0.011 & 0.442±0.004 & \textbf{0.394±0.008} & 0.450±0.010 & 0.375±0.007 & 0.300±0.007 & 0.281±0.004 & 0.225±0.002 & 0.195±0.011 \\
 &  & QWEN3-VL & 0.450±0.011 & 0.358±0.008 & 0.328±0.005 & 0.368±0.005 & 0.263±0.009 & 0.246±0.011 & 0.079±0.011 & 0.118±0.010 & 0.124±0.008 \\
\bottomrule
\end{tabular}

}
\caption{%
\textbf{Full item recommendation results across all scenarios.}
We report \textit{Precision@}1, \textit{Precision@}2, and \textit{Precision@}3 for all evaluated item recommendation methods, averaged over all queries within each scenario.
Values are shown with 95\% confidence intervals computed via bootstrap resampling.
Higher values indicate better alignment between recommended items and the ground-truth items inferred from the conversation prefix.}
\label{tab:item_rec}
\end{table*}

\paragraph{Movie.} The prompt used for experiments in Movie dataset is

\begin{lstlisting}[language={}, basicstyle=\ttfamily\scriptsize, breaklines=true]
Important Constraint:
Do NOT select attributes that are visually present or easily inferred from posters or thumbnails.

Examples of visually inferable attributes in Movie:
1. "The movie title and release year are shown on the poster."
2. "The poster highlights the lead actors prominently."
3. "The visual style clearly suggests a superhero action genre."

Assume the user can already perceive titles, actor names, and high-level visual cues.

Focus instead on attributes that clarify narrative structure, pacing,
emotional intensity, themes, or viewing suitability.
\end{lstlisting}

\paragraph{Retail .} The prompt used for experiments in Retail  dataset is

\begin{lstlisting}[language={}, basicstyle=\ttfamily\scriptsize, breaklines=true]
Important Constraint  :
Do NOT select attributes that are immediately observable from the product.

Examples of visually inferable attributes in Retail:
1. "The product is compact and rectangular in shape."
2. "The packaging is black with a glossy finish."
3. "The brand logo is clearly visible on the item."

Assume the user can see the product's size, shape, color, and packaging.

Focus instead on attributes related to functionality, reliability,
compatibility, warranty, or maintenance.
\end{lstlisting}

\section{Conversation Prefix Comparison in Item Recommendation}
\label{app:conversation_prefix}

We further evaluate the robustness of item recommendation performance under different choices of conversation prefixes. In particular, we compare the default prefix, defined as the conversation history prior to the first ground-truth recommendation, with an alternative setting that incorporates the \emph{full conversation history}, which represents the maximum information available under the standard definition of CRS.

To prevent trivial item identification through explicit mentions of ground-truth item names in the full conversation, we mask all utterances that explicitly reference recommended item names. This masking ensures that the evaluation does not degenerate into simple string matching, thereby preserving the intended objective of item recommendation in CRS.

As shown in \Autoref{full_cov}, we evaluate item recommendation performance across three scenarios using the methods presented in the main manuscript. For the Movie and Retail scenarios, incorporating the full conversation history does not yield significant performance gains. In contrast, for the Fashion scenario, all methods exhibit substantial improvements when provided with the full conversation.

We hypothesize that this gain arises because, in the Fashion scenario, seekers and recommenders frequently discuss detailed attributes of the ground-truth item \emph{after} the recommendation is made to support decision-making. When the full conversation is used as input, these post-recommendation attribute descriptions can uniquely identify the ground-truth items, leading to unintended information leakage.

To avoid this leakage and ensure a fair evaluation of item recommendation, we adopt the conversation prefix that ends before any ground-truth items are explicitly mentioned. Finally, we do not evaluate label selection under the full-conversation setting, as the full conversation contains utterances annotated with intent tags that are used to construct the label-selection ground truth.

\section{Full Evaluation Results}
\label{app:full_result}

This appendix reports the complete quantitative results for both \emph{Item Recommendation} and \emph{Immersive Label Selection} across all evaluated methods and scenarios.

Key findings from these results are discussed in Section 5. Briefly, across all criteria, GPT-5.1 consistently achieves the strongest performance on RIN (Table 3), while all methods show substantial degradation on EIN (Table 4) and PIN (Table 5), with several methods approaching the random baseline on PIN in the Movie scenario. Confidence intervals are uniformly narrow for RIN but widen for PIN, reflecting the increased difficulty and variance of proactive need anticipation.

\paragraph{Item Recommendation.}
\Autoref{tab:item_rec} presents full Item Recommendation results, reporting \texttt{Precision@1}, \texttt{Precision@2}, and \texttt{Precision@3}, averaged over all queries within each scenario. We additionally report 95\% confidence intervals for all metrics to reflect statistical variability.

\paragraph{Label Selection.}
Full results for immersive label selection are summarized in \Autoref{tab:eis}, \Autoref{tab:ine}, and \Autoref{tab:ins}, corresponding respectively to the \textit{RIN}, \textit{EIN}, and \textit{PIN} evaluation criteria. For each criterion, we report \texttt{mP@1}, \texttt{mP@2}, and \texttt{mP@3} across all evaluated methods.

\begin{table*}
\renewcommand{\arraystretch}{1.2}
\scriptsize
\centering
\resizebox{\linewidth}{!}{
    \begin{tabular}{lll|lll|lll|lll}
\hline
 &  & data & \multicolumn{3}{c}{\textbf{Fashion}} & \multicolumn{3}{c}{\textbf{Retail}} & \multicolumn{3}{c}{\textbf{Movie}} \\

 &  & metric & Precision@1 & Precision@2 & Precision@3 & Precision@1 & Precision@2 & Precision@3 & Precision@1 & Precision@2 & Precision@3 \\
method & form & model &  &  &  &  &  &  &  &  &  \\
\hline
BM25 &  &  & 0.425 ± 0.023 & 0.421 ± 0.018 & 0.404 ± 0.016 & 0.500 ± 0.044 & 0.475 ± 0.029 & 0.458 ± 0.025 & 0.221 ± 0.022 & 0.262 ± 0.018 & 0.264 ± 0.017 \\
\hline
CLIP &  & CLIP-LARGE & 0.702 ± 0.024 & 0.666 ± 0.020 & 0.638 ± 0.017 & 0.300 ± 0.038 & 0.425 ± 0.034 & 0.375 ± 0.022 & 0.651 ± 0.026 & 0.599 ± 0.020 & 0.554 ± 0.017 \\
\hline
CROSS ENCODE &  & miniLM-v6 & 0.437 ± 0.027 & 0.399 ± 0.018 & 0.366 ± 0.016 & 0.500 ± 0.041 & 0.388 ± 0.035 & 0.392 ± 0.034 & 0.372 ± 0.026 & 0.314 ± 0.019 & 0.298 ± 0.018 \\
\hline
Dense &  & QWEN3-8B & 0.675 ± 0.023 & 0.680 ± 0.017 & 0.656 ± 0.015 & 0.368 ± 0.046 & 0.408 ± 0.027 & 0.421 ± 0.028 & 0.453 ± 0.027 & 0.500 ± 0.018 & 0.481 ± 0.015 \\
\hline
\multirow[t]{6}{*}{LLM (T)} & \multirow[t]{3}{*}{listwise} & GEMINI-2.5P & 0.590 ± 0.025 & 0.581 ± 0.017 & 0.549 ± 0.017 & 0.725 ± 0.038 & 0.688 ± 0.030 & 0.642 ± 0.029 & 0.419 ± 0.027 & 0.419 ± 0.022 & 0.353 ± 0.019 \\
 &  & GPT-5.1 & 0.899 ± 0.014 & 0.797 ± 0.013 & 0.728 ± 0.015 & \textbf{0.775 ± 0.038} & 0.700 ± 0.032 & 0.625 ± 0.026 & 0.779 ± 0.022 & 0.570 ± 0.019 & 0.477 ± 0.014 \\
 &  & QWEN3-VL & 0.692 ± 0.023 & 0.642 ± 0.018 & 0.603 ± 0.016 & 0.525 ± 0.042 & 0.463 ± 0.033 & 0.400 ± 0.025 & 0.430 ± 0.027 & 0.453 ± 0.019 & 0.442 ± 0.017 \\
\cline{2-12}
 & \multirow[t]{3}{*}{pointwise} & GEMINI-2.5P & 0.771 ± 0.021 & 0.760 ± 0.019 & 0.729 ± 0.018 & 0.650 ± 0.045 & 0.637 ± 0.023 & 0.642 ± 0.018 & 0.477 ± 0.027 & 0.541 ± 0.017 & 0.488 ± 0.014 \\
 &  & GPT-5.1 & \textbf{0.907 ± 0.013} & \textbf{0.886 ± 0.011} & \textbf{0.862 ± 0.013} & 0.750 ± 0.034 & \textbf{0.762 ± 0.026} & 0.700 ± 0.022 & 0.593 ± 0.027 & 0.669 ± 0.016 & 0.605 ± 0.015 \\
 &  & QWEN3-VL & 0.704 ± 0.021 & 0.627 ± 0.019 & 0.615 ± 0.019 & 0.525 ± 0.042 & 0.412 ± 0.028 & 0.392 ± 0.024 & 0.198 ± 0.022 & 0.285 ± 0.018 & 0.360 ± 0.015 \\
\hline
\multirow[t]{6}{*}{VLM (V + T)} & \multirow[t]{3}{*}{listwise} & GEMINI-2.5P & 0.469 ± 0.025 & 0.461 ± 0.019 & 0.449 ± 0.017 & 0.625 ± 0.036 & 0.562 ± 0.036 & 0.517 ± 0.033 & 0.291 ± 0.025 & 0.250 ± 0.020 & 0.244 ± 0.017 \\
 &  & GPT-5.1 & 0.905 ± 0.016 & 0.749 ± 0.015 & 0.685 ± 0.015 & \textbf{0.775 ± 0.038} & 0.662 ± 0.032 & 0.583 ± 0.030 & \textbf{0.814 ± 0.021} & 0.581 ± 0.017 & 0.473 ± 0.014 \\
 &  & QWEN3-VL & 0.647 ± 0.024 & 0.629 ± 0.018 & 0.586 ± 0.016 & 0.500 ± 0.041 & 0.487 ± 0.031 & 0.442 ± 0.029 & 0.314 ± 0.025 & 0.384 ± 0.018 & 0.380 ± 0.015 \\
\cline{2-12}
 & \multirow[t]{3}{*}{pointwise} & GEMINI-2.5P & 0.776 ± 0.020 & 0.764 ± 0.018 & 0.720 ± 0.017 & 0.750 ± 0.034 & 0.625 ± 0.025 & 0.617 ± 0.019 & 0.547 ± 0.027 & 0.581 ± 0.016 & 0.523 ± 0.014 \\
 &  & GPT-5.1 & 0.896 ± 0.014 & 0.882 ± 0.013 & 0.861 ± 0.013 & 0.750 ± 0.029 & 0.738 ± 0.028 & \textbf{0.733 ± 0.024} & 0.709 ± 0.025 & \textbf{0.686 ± 0.017} & \textbf{0.624 ± 0.015} \\
 &  & QWEN3-VL & 0.626 ± 0.027 & 0.547 ± 0.021 & 0.541 ± 0.019 & 0.450 ± 0.048 & 0.362 ± 0.033 & 0.358 ± 0.028 & 0.058 ± 0.013 & 0.203 ± 0.013 & 0.267 ± 0.012 \\
\hline
RANDOM &  &  & 0.121 ± 0.017 & 0.322 ± 0.015 & 0.366 ± 0.014 & 0.125 ± 0.025 & 0.200 ± 0.023 & 0.242 ± 0.023 & 0.047 ± 0.011 & 0.058 ± 0.009 & 0.054 ± 0.007 \\
\hline
RERANK &  &  & 0.664 ± 0.022 & 0.583 ± 0.017 & 0.569 ± 0.015 & 0.550 ± 0.040 & 0.525 ± 0.027 & 0.500 ± 0.026 & 0.512 ± 0.027 & 0.430 ± 0.021 & 0.407 ± 0.018 \\
\hline
\bottomrule
\end{tabular}

}
\caption{%
\textbf{Label selection results under the \textit{RIN} criterion.}
Mean Precision at $k$ (\textit{mP@}1, 2, 3) for immersive label selection evaluated against the \emph{Retroactive Information Need (\textit{RIN})} ground truth.
This criterion measures whether selected labels correctly address information needs that are explicitly stated in the conversation.
All results are reported with 95\% confidence intervals. See Section 5.2 for discussion of key trends.}
\label{tab:eis}
\end{table*}

\begin{table*}
\renewcommand{\arraystretch}{1.2}
\scriptsize
\centering
\resizebox{\linewidth}{!}{
    \begin{tabular}{lll|lll|lll|lll}
\hline
 &  & data & \multicolumn{3}{c}{\textbf{Fashion}} & \multicolumn{3}{c}{\textbf{Retail}} & \multicolumn{3}{c}{\textbf{Movie}} \\

 &  & metric & Precision@1 & Precision@2 & Precision@3 & Precision@1 & Precision@2 & Precision@3 & Precision@1 & Precision@2 & Precision@3 \\
method & form & model &  &  &  &  &  &  &  &  &  \\
\hline
BM25 &  &  & 0.383 ± 0.024 & 0.381 ± 0.019 & 0.371 ± 0.016 & 0.353 ± 0.036 & 0.353 ± 0.028 & 0.304 ± 0.021 & 0.250 ± 0.024 & 0.281 ± 0.020 & 0.308 ± 0.017 \\
\hline
CLIP &  & CLIP-LARGE & 0.498 ± 0.025 & 0.479 ± 0.021 & 0.453 ± 0.019 & 0.294 ± 0.043 & 0.294 ± 0.033 & 0.255 ± 0.029 & 0.287 ± 0.025 & 0.206 ± 0.018 & 0.204 ± 0.015 \\
\hline
CROSS ENCODE &  & miniLM-v6 & 0.378 ± 0.025 & 0.319 ± 0.018 & 0.312 ± 0.017 & 0.324 ± 0.043 & 0.265 ± 0.038 & 0.275 ± 0.034 & 0.300 ± 0.026 & 0.287 ± 0.020 & 0.279 ± 0.018 \\
\hline
Dense &  & QWEN3-8B & 0.569 ± 0.025 & 0.508 ± 0.019 & 0.462 ± 0.017 & 0.312 ± 0.045 & 0.281 ± 0.025 & 0.292 ± 0.027 & 0.163 ± 0.021 & 0.150 ± 0.016 & 0.154 ± 0.013 \\
\hline
\multirow[t]{6}{*}{VLM (T)} & \multirow[t]{3}{*}{listwise} & GEMINI-2.5P & 0.320 ± 0.024 & 0.297 ± 0.017 & 0.323 ± 0.014 & 0.235 ± 0.044 & 0.324 ± 0.028 & 0.294 ± 0.023 & 0.188 ± 0.022 & 0.188 ± 0.016 & 0.208 ± 0.016 \\
 &  & gpt-5.1 & 0.588 ± 0.023 & 0.532 ± 0.020 & \textbf{0.532 ± 0.017} & \textbf{0.529 ± 0.050} & \textbf{0.426 ± 0.040} & \textbf{0.392 ± 0.025} & \textbf{0.500 ± 0.028} & \textbf{0.419 ± 0.022} & 0.354 ± 0.018 \\
 &  & QWEN3-VL & 0.501 ± 0.024 & 0.492 ± 0.019 & 0.487 ± 0.017 & 0.294 ± 0.037 & 0.250 ± 0.019 & 0.245 ± 0.018 & 0.388 ± 0.027 & 0.325 ± 0.018 & 0.296 ± 0.016 \\
\cline{2-12}
 & \multirow[t]{3}{*}{pointwise} & GEMINI-2.5P & 0.500 ± 0.026 & 0.497 ± 0.022 & 0.461 ± 0.018 & 0.353 ± 0.047 & 0.324 ± 0.032 & 0.314 ± 0.026 & 0.037 ± 0.011 & 0.212 ± 0.015 & 0.254 ± 0.015 \\
 &  & gpt-5.1 & 0.546 ± 0.024 & 0.518 ± 0.021 & 0.523 ± 0.019 & 0.294 ± 0.037 & 0.353 ± 0.028 & 0.382 ± 0.022 & 0.237 ± 0.024 & 0.356 ± 0.020 & 0.388 ± 0.019 \\
 &  & QWEN3-VL & 0.462 ± 0.023 & 0.443 ± 0.019 & 0.437 ± 0.018 & 0.265 ± 0.031 & 0.176 ± 0.026 & 0.167 ± 0.026 & 0.062 ± 0.014 & 0.094 ± 0.013 & 0.154 ± 0.013 \\
\hline
\multirow[t]{6}{*}{VLM (V + T)} & \multirow[t]{3}{*}{listwise} & GEMINI-2.5P & 0.288 ± 0.023 & 0.295 ± 0.018 & 0.312 ± 0.014 & 0.265 ± 0.031 & 0.294 ± 0.027 & 0.284 ± 0.020 & 0.200 ± 0.023 & 0.194 ± 0.018 & 0.171 ± 0.014 \\
 &  & gpt-5.1 & \textbf{0.605 ± 0.028} & \textbf{0.543 ± 0.022} & 0.507 ± 0.019 & 0.441 ± 0.042 & 0.412 ± 0.032 & 0.373 ± 0.026 & \textbf{0.500 ± 0.028} & 0.388 ± 0.021 & 0.346 ± 0.018 \\
 &  & QWEN3-VL & 0.508 ± 0.024 & 0.472 ± 0.019 & 0.461 ± 0.017 & 0.265 ± 0.038 & 0.279 ± 0.021 & 0.255 ± 0.022 & 0.388 ± 0.027 & 0.344 ± 0.020 & 0.296 ± 0.018 \\
\cline{2-12}
 & \multirow[t]{3}{*}{pointwise} & GEMINI-2.5P & 0.486 ± 0.027 & 0.514 ± 0.022 & 0.485 ± 0.019 & 0.324 ± 0.037 & 0.294 ± 0.029 & 0.284 ± 0.023 & 0.138 ± 0.019 & 0.269 ± 0.018 & 0.287 ± 0.017 \\
 &  & gpt-5.1 & 0.562 ± 0.025 & 0.519 ± 0.021 & 0.531 ± 0.019 & 0.382 ± 0.034 & 0.382 ± 0.024 & 0.373 ± 0.024 & 0.225 ± 0.023 & 0.344 ± 0.020 & \textbf{0.400 ± 0.020} \\
 &  & QWEN3-VL & 0.422 ± 0.027 & 0.430 ± 0.022 & 0.409 ± 0.017 & 0.324 ± 0.037 & 0.235 ± 0.027 & 0.235 ± 0.024 & 0.038 ± 0.011 & 0.082 ± 0.012 & 0.156 ± 0.012 \\
\hline
RANDOM &  &  & 0.111 ± 0.016 & 0.234 ± 0.016 & 0.300 ± 0.015 & 0.206 ± 0.031 & 0.235 ± 0.027 & 0.265 ± 0.024 & 0.175 ± 0.021 & 0.156 ± 0.014 & 0.137 ± 0.011 \\
\hline
RERANK &  &  & 0.466 ± 0.026 & 0.441 ± 0.019 & 0.457 ± 0.018 & 0.382 ± 0.046 & 0.353 ± 0.032 & 0.333 ± 0.021 & 0.263 ± 0.025 & 0.256 ± 0.019 & 0.242 ± 0.016 \\
\bottomrule
\end{tabular}

}
\caption{%
\textbf{Label selection results under the \textit{EIN} criterion.}
Mean Precision at $k$ (\textit{mP@}1, 2, 3) evaluated against the \emph{Expert Inferred Information Need (EIN)} ground truth.
This criterion captures alignment with expert-annotated latent information needs that are not visually inferable from the item visual information but are relevant for decision support.
All values include 95\% confidence intervals. See Section 5.2 for discussion of key trends.}
\label{tab:ine}
\end{table*}

\begin{table*}
\renewcommand{\arraystretch}{1.2}
\scriptsize
\centering
\resizebox{\linewidth}{!}{
    \begin{tabular}{lll|lll|lll|lll}
\hline
 &  & data & \multicolumn{3}{c}{\textbf{Fashion}} & \multicolumn{3}{c}{\textbf{Retail}} & \multicolumn{3}{c}{\textbf{Movie}} \\

 &  & metric & Precision@1 & Precision@2 & Precision@3 & Precision@1 & Precision@2 & Precision@3 & Precision@1 & Precision@2 & Precision@3 \\
method & form & model &  &  &  &  &  &  &  &  &  \\
\hline
BM25 &  &  & 0.244 ± 0.034 & 0.186 ± 0.027 & 0.179 ± 0.021 & 0.367 ± 0.052 & 0.450 ± 0.044 & 0.378 ± 0.043 & 0.297 ± 0.027 & \textbf{0.277 ± 0.022} & \textbf{0.288 ± 0.019} \\
\hline
CLIP &  & CLIP-LARGE & 0.282 ± 0.037 & 0.288 ± 0.032 & 0.273 ± 0.026 & 0.133 ± 0.038 & 0.200 ± 0.030 & 0.189 ± 0.024 & 0.081 ± 0.016 & 0.054 ± 0.009 & 0.068 ± 0.009 \\
\hline
CROSS ENCODE &  & miniLM-v6 & 0.058 ± 0.021 & 0.115 ± 0.025 & 0.124 ± 0.025 & 0.267 ± 0.041 & 0.200 ± 0.033 & 0.189 ± 0.031 & 0.162 ± 0.022 & 0.182 ± 0.018 & 0.207 ± 0.017 \\
\hline
Dense &  & QWEN3-8B & 0.205 ± 0.031 & 0.221 ± 0.022 & 0.209 ± 0.022 & 0.143 ± 0.031 & 0.179 ± 0.024 & 0.190 ± 0.026 & 0.081 ± 0.016 & 0.054 ± 0.009 & 0.063 ± 0.008 \\
\hline
\multirow[t]{6}{*}{VLM (T)} & \multirow[t]{3}{*}{listwise} & GEMINI-2.5P & 0.186 ± 0.027 & 0.199 ± 0.020 & 0.179 ± 0.014 & 0.267 ± 0.048 & 0.333 ± 0.043 & 0.300 ± 0.035 & 0.189 ± 0.023 & 0.203 ± 0.017 & 0.194 ± 0.014 \\
 &  & gpt-5.1 & 0.385 ± 0.042 & 0.343 ± 0.037 & 0.325 ± 0.029 & \textbf{0.600 ± 0.056} & 0.450 ± 0.039 & 0.389 ± 0.031 & 0.189 ± 0.023 & 0.236 ± 0.019 & 0.203 ± 0.016 \\
 &  & QWEN3-VL & 0.429 ± 0.041 & 0.365 ± 0.037 & 0.346 ± 0.030 & 0.267 ± 0.048 & 0.200 ± 0.033 & 0.167 ± 0.024 & \textbf{0.338 ± 0.028} & 0.257 ± 0.019 & 0.230 ± 0.017 \\
\cline{2-12}
 & \multirow[t]{3}{*}{pointwise} & GEMINI-2.5P & 0.276 ± 0.040 & 0.301 ± 0.036 & 0.299 ± 0.032 & 0.533 ± 0.057 & 0.433 ± 0.046 & 0.378 ± 0.040 & 0.014 ± 0.007 & 0.047 ± 0.009 & 0.108 ± 0.011 \\
 &  & gpt-5.1 & 0.365 ± 0.043 & \textbf{0.404 ± 0.037} & \textbf{0.365 ± 0.033} & 0.500 ± 0.060 & 0.367 ± 0.047 & 0.400 ± 0.043 & 0.095 ± 0.017 & 0.155 ± 0.016 & 0.207 ± 0.016 \\
 &  & QWEN3-VL & 0.314 ± 0.037 & 0.266 ± 0.031 & 0.261 ± 0.030 & 0.267 ± 0.054 & 0.200 ± 0.041 & 0.178 ± 0.031 & 0.027 ± 0.009 & 0.047 ± 0.010 & 0.077 ± 0.011 \\
\hline
\multirow[t]{6}{*}{VLM (V + T)} & \multirow[t]{3}{*}{listwise} & GEMINI-2.5P & 0.192 ± 0.028 & 0.173 ± 0.021 & 0.160 ± 0.015 & 0.333 ± 0.053 & 0.300 ± 0.041 & 0.289 ± 0.035 & 0.162 ± 0.022 & 0.182 ± 0.018 & 0.185 ± 0.016 \\
 &  & gpt-5.1 & \textbf{0.520 ± 0.044} & 0.363 ± 0.033 & 0.293 ± 0.024 & 0.433 ± 0.054 & 0.417 ± 0.038 & 0.378 ± 0.034 & 0.216 ± 0.024 & 0.203 ± 0.017 & 0.203 ± 0.015 \\
 &  & QWEN3-VL & 0.420 ± 0.045 & 0.367 ± 0.037 & 0.331 ± 0.030 & 0.267 ± 0.041 & 0.217 ± 0.034 & 0.189 ± 0.028 & 0.311 ± 0.027 & 0.264 ± 0.019 & 0.239 ± 0.016 \\
\cline{2-12}
 & \multirow[t]{3}{*}{pointwise} & GEMINI-2.5P & 0.205 ± 0.037 & 0.279 ± 0.030 & 0.276 ± 0.028 & 0.500 ± 0.055 & \textbf{0.467 ± 0.049} & \textbf{0.422 ± 0.042} & 0.041 ± 0.012 & 0.074 ± 0.011 & 0.149 ± 0.013 \\
 &  & gpt-5.1 & 0.380 ± 0.044 & 0.370 ± 0.040 & 0.360 ± 0.035 & 0.533 ± 0.062 & 0.417 ± 0.043 & 0.411 ± 0.039 & 0.041 ± 0.012 & 0.108 ± 0.013 & 0.189 ± 0.014 \\
 &  & QWEN3-VL & 0.253 ± 0.038 & 0.230 ± 0.029 & 0.260 ± 0.025 & 0.433 ± 0.059 & 0.317 ± 0.043 & 0.244 ± 0.033 & 0.000 ± 0.000 & 0.041 ± 0.008 & 0.073 ± 0.010 \\
\hline
RANDOM &  &  & 0.020 ± 0.010 & 0.150 ± 0.020 & 0.168 ± 0.019 & 0.233 ± 0.041 & 0.317 ± 0.039 & 0.300 ± 0.033 & 0.095 ± 0.017 & 0.142 ± 0.016 & 0.149 ± 0.013 \\
\hline
RERANK &  &  & 0.333 ± 0.042 & 0.279 ± 0.032 & 0.241 ± 0.023 & 0.367 ± 0.062 & 0.383 ± 0.034 & 0.367 ± 0.033 & 0.162 ± 0.022 & 0.216 ± 0.019 & 0.207 ± 0.016 \\
\bottomrule
\end{tabular}

}
\caption{%
\textbf{Label selection results under the \textit{PIN} criterion.}
Mean Precision at $k$ (\textit{mP@}1, 2, 3) evaluated against the \emph{Proactive Information Need (PIN)} ground truth.
This criterion assesses whether selected labels align with the information that a seeker is likely to proactively request in subsequent turns to resolve uncertainty and support decision-making.
All results are reported with 95\% confidence intervals. See Section 5.2 for discussion of key trends.}
\label{tab:ins}
\end{table*}






\section{Pointwise VLM as Immersive Label Selection}
\label{app:pointwise}
\begin{figure}
    \centering
    \includegraphics[width=\linewidth]{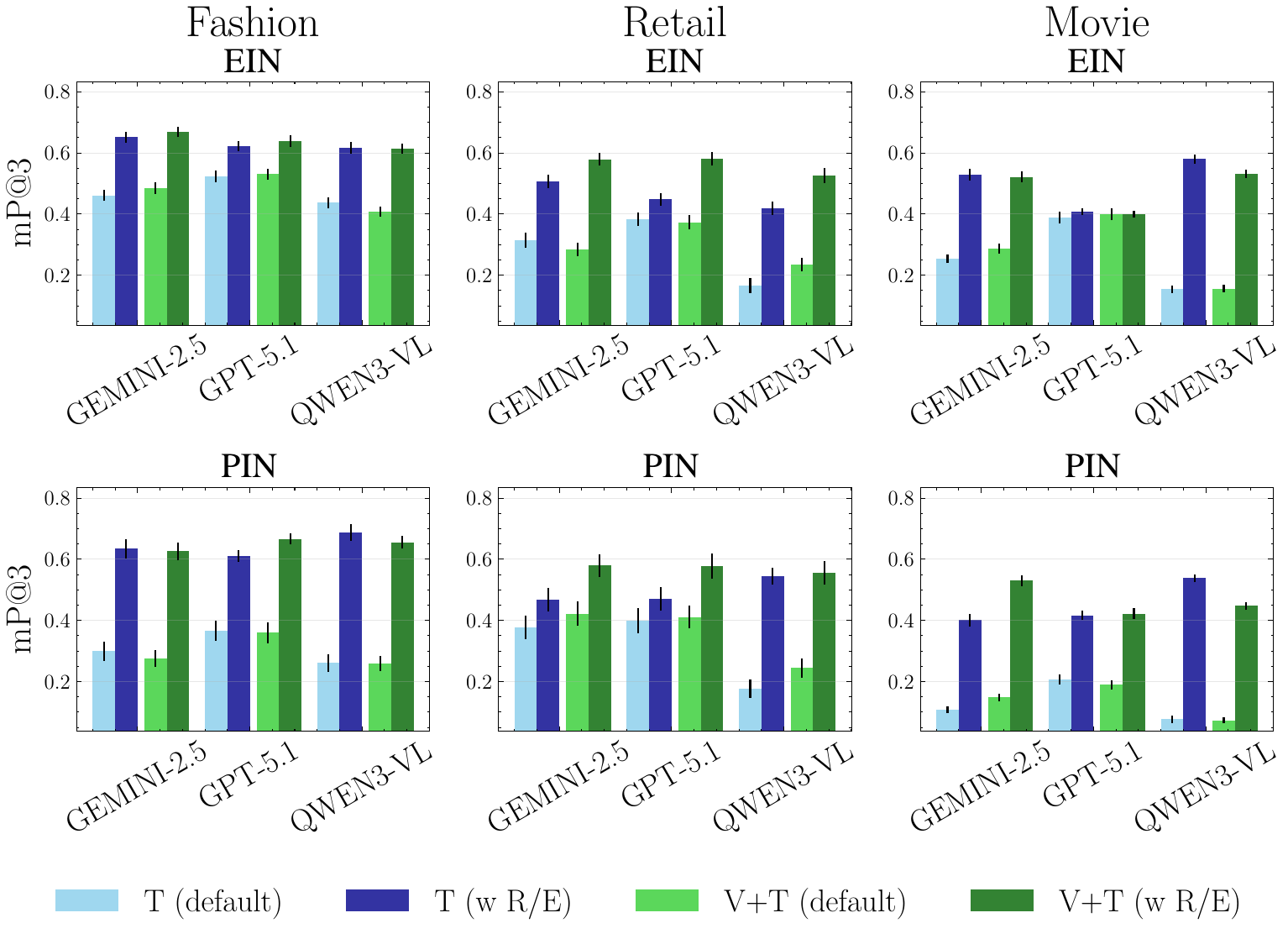}
\caption{pointwise VLM-based methods given utterances tagged as \textit{Implicit Seeker Request} and \textit{Expert Explanation} (\textsc{w R/E}), which simplifies inferring proactive needs into matching explicit requests.}
    \label{fig:app_pointwise-VLM_relax}
\end{figure}

\begin{figure}
    \centering
    \includegraphics[width=\linewidth]{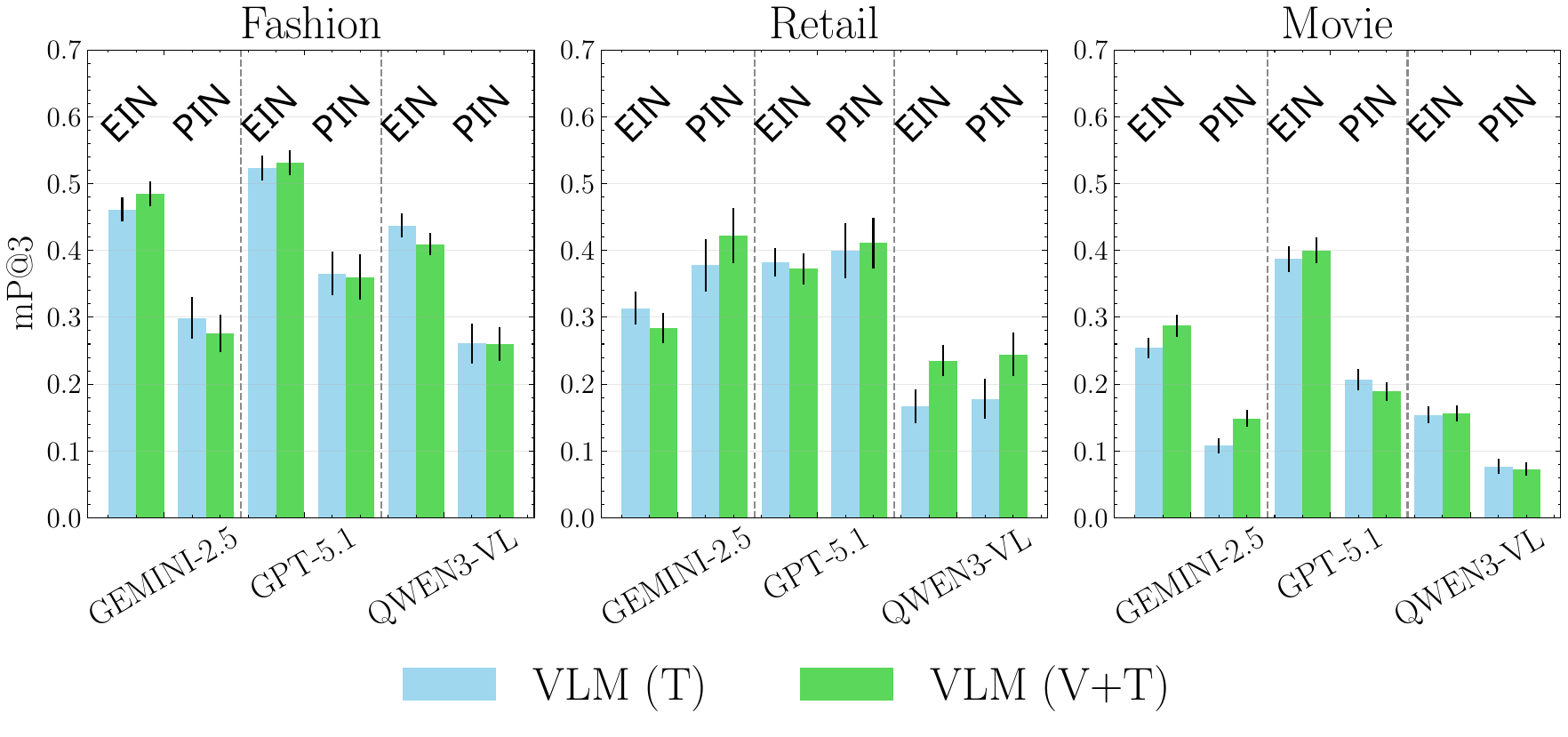}
        \includegraphics[width=\linewidth]{appendix_plots/llm_comparison_all_data_v_vt_P_3_p.pdf} 
\caption{(\textbf{Upper}) Pointwise VLM performance (\textsc{T}) versus (\textsc{V+T}); (\textbf{Lower}) Pointwise zero-shot VLM with few-shot instruction discouraging visually inferable attributes (\textsc{Instruct}) under \textsc{V+T}s.}
    \label{fig:app_pointwise-VLM_insturct}
\end{figure}

\begin{figure*}
    \centering
\includegraphics[width=0.8\linewidth]{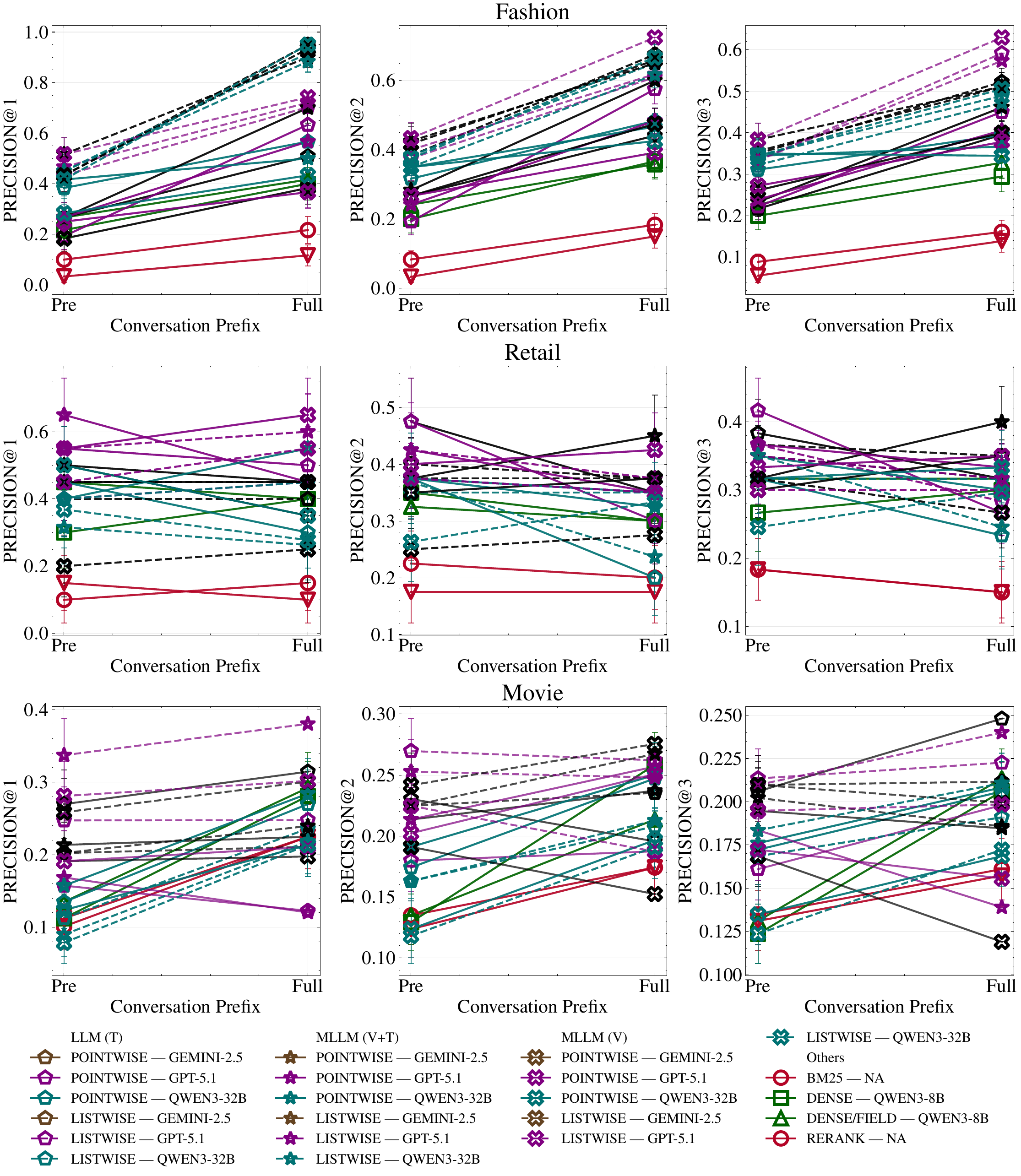}
\caption{\texttt{Precision@3} of using conversation before mentioning the first ground truth as the conversation prefix, which is used across experiments, vs using the full conversation history as the conversation prefix by masking the ground truth item's names}
\label{full_cov}
\end{figure*}

\Autoref{fig:app_pointwise-VLM_insturct} and \Autoref{fig:app_pointwise-VLM_relax} present a detailed analysis of pointwise zero-shot VLM methods for immersive label selection, complementing the main results by isolating the effects of visual grounding, instruction prompting, and conversation constraints. All results are reported using \textit{mP@3}.

Results largely mirror the listwise findings discussed in Section 5.2: pointwise methods achieve reasonable RIN performance but fail to anticipate proactive information needs (PIN), and few-shot instruction prompting (Figure 14, bottom) improves latent awareness primarily in the Movie scenario where visually inferable attributes are most prevalent.

\end{document}